\documentclass[a4paper,11pt]{article}
\pdfoutput=1 
\usepackage{jheppub} 
\usepackage{hyperref}
\usepackage{booktabs}
\usepackage[T1]{fontenc} 
\usepackage{siunitx} 
\usepackage{graphicx}%
\usepackage[dvipsnames]{xcolor}
\usepackage{dcolumn}
\usepackage{bm}
\usepackage[normalem]{ulem}
\usepackage{braket} 
\usepackage{amsmath}
\usepackage{cancel}
\usepackage{slashed}
\usepackage{multicol}
\usepackage[capitalise]{cleveref}

\counterwithin*{equation}{section}
\allowdisplaybreaks

\newcommand{\nnl}{\nonumber \\}

\newcommand{\cL}{\mathcal{L}}
\newcommand{\cO}{\mathcal{O}}

\newcommand{\cM}{{\mathcal M}}

\newcommand{\mpl}{M_{\rm Pl}}

\usepackage{etoolbox,siunitx}
\robustify\bfseries

\usepackage{subcaption}
\newcommand{\la}{\langle}
\newcommand{\ra}{\rangle}
\newcommand{\mla}[1]{\bold{\langle#1}}
\newcommand{\mra}[1]{\bold#1\rangle}
\newcommand{\mls}[1]{\bold{[#1}}
\newcommand{\mrs}[1]{\bold{#1]}}

\begin{document}

\title{
Matching and positivity beyond minimal coupling
}

\author{Edoardo Alviani and}
\author{Adam Falkowski }

\affiliation{Universit\'{e} Paris-Saclay, CNRS/IN2P3, IJCLab, 91405 Orsay, France}

\emailAdd{
alviani@ijclab.in2p3.fr,
adam.falkowski@ijclab.in2p3.fr
}

\abstract{
We study the low-energy effective theories of photons and gravitons, matching them at one loop to a UV completion with a massive spinning matter particle. 
The matter is allowed to have non-minimal  electromagnetic and gravitational interactions. 
We construct the one-loop 4-photon and 4-graviton helicity amplitudes with matter in the loop carrying anomalous multipole moments, and we read off the Wilson coefficients in the effective theory below the matter particle mass.   
Much as in the case of minimal coupling, the Wilson coefficients beyond the leading order turn out to be confined to small islands in the much larger theory space allowed by existing positivity constraints.  
}
\maketitle

\section{Introduction} 

Effective Field Theory~(EFT) is a framework underlying much of modern physics. 
The basic idea is to focus on large-scale behavior where the heavy degrees of freedom, not directly available in experiment, are integrated out. 
Their effects are encoded in the Wilson coefficients of a series of interaction terms of the light degrees of freedom, which are organized in a systematic expansion according to some power counting. 
The Wilson coefficients can be calculated via the matching procedure, when the UV completion of the EFT is known and is perturbative. 

It is often advantageous to use EFT together with on-shell amplitude techniques. 
The latter 
allow one to perform practical calculations, arguably in a simpler and more transparent fashion than the traditional techniques of quantum field theory.
In this approach, the basic building blocks are 3-particle on-shell amplitudes, which are very constrained by the requirements of Poincar\'{e} invariance, locality, and little group covariance.  
Higher-point amplitudes can be bootstrapped from these objects by making use of unitarity and locality. 
Fruitful applications of on-shell techniques include construction of EFT bases~\cite{Durieux:2019siw,Li:2020gnx,Durieux:2020gip,Dong:2021vxo,Li:2022tec,DeAngelis:2022qco} or calculation of renormalization group (RG) equations for the Wilson coefficients~\cite{Caron-Huot:2016cwu,Baratella:2020lzz,Bern:2020ikv,EliasMiro:2020tdv,Baratella:2021guc}.
The superiority of the on-shell approach is especially striking when the EFT involves particles with higher spin, such as e.g. the spin-2 graviton.

Two distinct EFTs (though sharing many similar features) will be the focus of this article.  Both are organized around the principles of local symmetry and power counting based on dimensional expansion. 
One, the so-called Euler-Heisenberg (EH) EFT~\cite{Heisenberg:1936nmg}, provides a low-energy description of spin-1 photons and its Lagrangian takes the form 
\begin{equation}
\label{eq:INT_Leheft}
\cL_{\rm EH} = 
- {1 \over 4} F_{\mu \nu} F_{\mu \nu}  
+  {C_1 \over 16}  (F_{\mu \nu} F_{\mu \nu})  (F_{\rho \sigma} F_{\rho \sigma})
+  {C_2 \over 16}   (F_{\mu \nu}  \tilde F_{\mu \nu})  (F_{\rho \sigma} \tilde F_{\rho \sigma})  
+ \dots 
\end{equation} 
where $F_{\mu \nu} = \partial_\mu A_\nu -  \partial_\nu A_\mu$, $\tilde F_{\mu \nu} = \epsilon_{\mu\nu\alpha \beta} F_{\alpha \beta}$,  and $A_\mu$ is a massless vector field representing the photon.
The Lagrangian is invariant under U(1) local transformations acting on the vector field. 
The lowest order photon self-interactions allowed by this symmetry occur at dimension 8. 
Power counting sets the corresponding Wilson coefficients to be of order $C_{1,2} \sim m^{-4}$, where $2 m$ is the cutoff of the EFT identified with the threshold for pair production of the lightest charged particle in the UV completion.\footnote{
If the Standard Model is the UV completion then $m$ is the electron mass, but we will allow for more general possibilities. }
The dots stand for operators of dimension 10 and higher, and their Wilson coefficients are suppressed by correspondingly higher powers of $m$. 
Here and in the following we assume parity conservation, for simplicity. 

The other EFT of interest here describes pure gravity, or equivalently self-interactions of a spin-2 particle called the graviton. 
The Lagrangian takes the form 
\begin{align}
\label{eq:INT_Lgreft}
\cL_{\rm GREFT} =   &  \sqrt{-g} \bigg \{ 
{\mpl^2  \over 2}  R 
+ {C_3 \over 3!}  R_{\mu \nu \alpha \beta} R_{\alpha \beta \rho \sigma}   R_{ \rho \sigma \mu \nu}
 \nnl + &
 {C_{4,1} \over 8}  (R_{\mu \nu \alpha \beta} R_{\mu \nu \alpha \beta})^2  
+ {C_{4,2} \over 8}   (R_{\mu \nu \alpha \beta} \tilde R_{\mu \nu \alpha \beta})^2  
 + \dots 
\bigg \}  
.   \end{align}  
Here, $R_{\mu \nu \alpha \beta}$ is the Riemann tensor built from the metric field $g_{\mu \nu}$ representing the graviton. 
The symmetry principle is general coordinate invariance, which is a local symmetry acting on the metric.  
The first term, with 
$\mpl \equiv (8 \pi G)^{-1/2} \simeq 2.43 \times 10^{18}$~GeV, 
is the familiar Einstein-Hilbert Lagrangian, encompassing the theory of general relativity (GR). 
Higher order interactions start at dimension six, with the interactions term cubic in the Riemann tensor, and the power counting 
$C_3 \sim \tilde C_3 \sim m^{-2}$. 
We also highlighted dimension-8 operators with 
$C_{4,1} \sim C_{4,2} \sim m^{-4}$. 
Once again the dots stand for interactions of dimension 10 and higher.  
This effective theory extension of GR is referred to as general relativity GREFT~\cite{Ruhdorfer:2019qmk}.

The interest in these particular EFTs is twofold. 
On one hand they are important tools to analyze a broad class of phenomena relevant for experiment. 
EHEFT captures physics of electromagnetic waves, from the longest wavelengths all the way to gamma rays. 
The dimension-8 interactions lead in particular to the phenomenon of vacuum birefringence (polarization-dependent change of propagation speed of electromagnetic waves in a magnetic field background), which has long been the target of many experiments~\cite{Ejlli:2020yhk}.
The dimension-6 and -8 interactions in GREFT affect in particular the black hole solutions in GR, leading to potentially observable effects in gravitational waves emission during inspirals and mergers~\cite{AccettulliHuber:2020dal,Brandhuber:2024bnz}.
On the other hand, there is purely theoretical interest too. 
While the EFT Wilson coefficients  may appear to be free parameters from the low-energy perspective,  they are in fact constrained in a subtle way given broad assumptions about the UV completion~\cite{Adams:2006sv,Bellazzini:2016xrt}, or often by demanding absence of superluminal propagation in non-trivial backgrounds~\cite{Camanho:2014apa,Cheung:2014ega,Bellazzini:2022wzv,CarrilloGonzalez:2022fwg}. 
This fact is often referred to as {\em positivity} constraints, 
and EHEFT or GREFT provide simple and interesting laboratories to study these.  
For example, the dimension-8 coefficients in \cref{eq:INT_Leheft} satisfy $C_{1(2)}>0$~\cite{brandotalk,Henriksson:2021ymi}, and similarly for the ones in~\cref{eq:INT_Lgreft}, $C_{4,1(2)}>0$~\cite{Bellazzini:2015cra}. 
Other coefficients may satisfy two-sided bounds~\cite{Tolley:2020gtv,Caron-Huot:2020cmc,Caron-Huot:2021rmr}, as is the case of $C_3$ in \cref{eq:INT_Lgreft}~\cite{Bern:2021ppb} and certain higher-order coefficients in \cref{eq:INT_Leheft}~\cite{Henriksson:2021ymi}. 
Determining the complete shape of the allowed theory space of these and other EFTs is subject to an ongoing and active research, see for example Refs.~\cite{Bellazzini:2020cot,Sinha:2020win,Arkani-Hamed:2020blm,Chiang:2021ziz,Caron-Huot:2022ugt,Davighi:2021osh,Haring:2022cyf}. 

From both of the perspectives mentioned above, it is interesting to connect the Wilson coefficients of EHEFT and GREFT to particular UV completions.
On the phenomenological side, this allows one to translate the  experimental constraints on the EFT Wilson coefficients  into  constraints on masses and coupling in the UV completions.
On the theoretical side, UV completions provide useful data to understand the EFT theory space. 
In particular, some UV completions may saturate the known bounds, proving that these bounds are already optimal. 
For example, the tree-level UV completion of EHEFT comprising a dilaton and an axion fills the entire positive quadrant of the $C_1$-$C_2$ plane. 
Conversely, it is also interesting when some regions of the parameter space allowed by positivity {\em cannot} be reached by known UV completions. 
This is the case for some higher-order coefficients in EHEFT and, even more spectacularly, for the GREFT Wilson coefficients. 
Indeed, it was argued in Ref.~\cite{Bern:2017puu} that known perturbative UV completions of gravity fill only small islands in the much vaster space consistent with existing positivity constraints. This suggests that stronger positivity constraints can be derived, or that unknown UV completions exist (a possibility particularly fascinating in the gravitational context), or both.  

In this paper we extend the calculation of EHEFT and GREFT Wilson coefficients to a larger class of UV completions. 
Namely, we integrate out at one loop particles up to spin $S=1(2)$ that are non-minimally coupled to photons (gravitons), with arbitrary multipole moments $\delta_n$, $n \leq 2S$. 
This generalizes the calculations in Refs.~\cite{Henriksson:2021ymi,Bern:2021ppb}, which assume minimal couplings.  
In the electromagnetic framework, this means we take UV completions to be EFT themselves, with the cutoff scale parametrically separated from particles' masses, $\Lambda \gg m$ and setting the magnitude of (anomalous) multipole corrections, $\delta_n \sim (m/\Lambda)^{2S}$. 
For gravity, the logic is similar, except in this case UV completions with a finite number of matter particles are EFTs themselves, with a finite validity range even for the minimal coupling. 
We study how these UV completions fill the parameter space of EFTs. 
For minimal coupling and a given spin and mass, integrating out a single particle gives a point in parameter space. 
In our case, multipole deformations allow one to migrate  that point in the theory space. 
Our focus is to quantify that migration, and compare it with existing positivity constraints. We use on-shell methods and spinor helicity formalism together with unitarity, which allows us to handle the computations efficiently, especially for higher spins.

This paper is organized as follows. 
In \cref{sec:COMPTON} we introduce the prerequisites for our calculation: the (gravi-)Compton amplitudes describing scattering of a massive matter particle with spin $S \leq 1(2)$ on a photon (graviton). 
Starting with a general on-shell 3-particle amplitude describing matter coupling to a photon or a graviton, and using the standard unitarity bootstrap techniques,  we obtain compact representations of leading multipole corrections to tree-level Compton amplitudes.

These amplitudes are an input for deriving one-loop amplitudes of interest using generalized unitarity.
First, 4-photon scattering amplitudes are derived in~\cref{sec:EH}. 
We lay out in some detail our method to calculate the box, triangle, and bubble coefficients as well as the rational terms of the 4-photon amplitude, and provide a simple power-counting argument why the tadpoles do not contribute for $S \leq 1$. 
Given the amplitude, it is straightforward to expand it in powers of $1/m$ and derive Wilson coefficients at arbitrary dimensions. 
This way we obtain linear multipole corrections to well known minimal couplings results for the Wilson coefficients. 
As we will explain, quadratic and higher-order corrections in multipoles are not always calculable within our EFT setup, and we give a succinct criterion for calculability.  
The analogous program is then carried out for  4-graviton scattering amplitudes in~\cref{sec:GREFT}. 
One additional subtlety here is that the tadpoles (and massless bubbles) do contribute, but it can be shown that the resulting ambiguity does not affect the Wilson coefficients we are interested in. 
Once again we expand the amplitudes in powers of $m$ to calculate $C_3$, $C_{4,1(2)}$, and higher-order Wilson coefficients with linear quadrupole corrections. 

In \cref{sec:positivity} we discuss how the results we have obtained fit into the known positivity constraints. 
In the EHEFT case, integrating out particles with anomalous dipole moments allows one to asymptotically approach the boundary
$C_2=0$ of the positive $C_1$-$C_2$ quadrant allowed by positivity, 
although this happens in the regime where unknwon corrections from particles above $\Lambda$ are expected to be significant.
For GREFT, the striking conclusion is that anomalous gravi-quadrupole moments practically  do not open any new parameter space compared to minimal coupling.   
Instead, due to a non-trivial conspiracy,  the Wilson coefficients remains in the small islands drawn up in Ref.~\cite{Bern:2021ppb}. 

Below we summarize the conventions. 
We  work with the mostly minus metric $\eta_{\mu \nu} = (1,-1,-1,-1)$, 
and the natural units $\hbar = c = 1$.
The sign convention for the totally anti-symmetric Levi-Civita tensor $\epsilon^{\mu\nu\rho\alpha}$ is 
$\epsilon^{0123} = 1$.    
The Christoffel connection is defined as $\Gamma^\mu_{\ \nu \rho} =   {1 \over 2}  g^{\mu \alpha} \left (\partial_\rho g_{\alpha \nu} +  \partial_\nu g_{\alpha \rho} - \partial_\alpha g_{\nu \rho} \right )$, 
the Riemann tensor is  $R^\alpha_{\ \mu \nu \beta}  =    \partial_\nu \Gamma^\alpha_{\ \mu \beta} -  \partial_\beta \Gamma^\alpha_{\ \mu \nu}    +  \Gamma^\rho_{\ \mu \beta}  \Gamma^\alpha_{\ \rho \nu}
 - \Gamma^\rho_{\ \mu \nu}  \Gamma^\alpha_{\ \rho \beta} $,  and the Ricci tensor is 
$R_{\mu \nu}  =  R^\alpha_{\ \mu \nu \alpha}$. 
We use the all-incoming convention for our on-shell amplitudes, unless otherwise noted.  
For helicity spinors, we use the conventions of Ref.~\cite{Dreiner:2008tw}, plus the usual shorthand notation: 
$\lambda_p^\alpha \equiv |p\ra$, 
$\tilde \lambda_p^{\dot \alpha} \equiv |p]$  
for massless 2-component spinors, 
$\alpha=1,2$, 
and 
$\chi_k^{\alpha\, J} \equiv |\mra{k}^J$, 
$\tilde \chi_k^{\dot \alpha\, J} \equiv |\mrs{k}^J$ for massive spinors, with $J=1,2$ encoding polarization.
In our conventions,
$\langle k l \rangle \equiv
\lambda_k^\alpha \lambda_{l\alpha}$,
$[ k l ]\equiv
\tilde \lambda_{k\dot \alpha} \tilde \lambda_l^{\dot \alpha}$, 
$\langle k | p | l ] \equiv
\lambda_k^\alpha [p \sigma]_{\alpha \dot \beta} \tilde \lambda_l^{\dot \beta}$, 
$[k | p | l \rangle \equiv
\tilde \lambda_{k\dot \alpha} [p \bar \sigma]^{\dot \alpha \beta} \tilde \lambda_{l\beta}$,
with 
$\sigma^\mu \equiv (1,\vec{\sigma})$,
$\bar \sigma^\mu \equiv (1,-\vec{\sigma})$.

\section{Compton amplitudes with multipoles} 
\label{sec:COMPTON}

As we will detail in the next section, 
we employ unitarity methods to match low-energy EFTs of photons and gravitons to more fundamental theories containing in addition matter particles.
The main ingredient for this calculation will be the Compton amplitudes. 
In this section we write down the (gravi-)Compton amplitudes describing scattering of a photon (graviton) on a matter particle $X$ of mass $m$ and spin $S\leq1$ ($S\leq2$). 
The matter is assumed to have non-minimal couplings to photons and gravitons through general multipole interactions. 
Such amplitudes can be computed starting from their residues in each kinematic channel and finding the unique amplitude constructable from those residues~\cite{Arkani-Hamed:2017jhn,Chung:2018kqs}.
Factorization properties required by unitarity 
allow us to bootstrap the residues  from the following set of on-shell three-point amplitudes~\cite{Arkani-Hamed:2017jhn}:
\begin{align}
\label{eq:CA_M3}
    &\cM\left (1_\gamma^- \bold2_X \bold3_{\bar X}\right)= - \sqrt{2} q_X  
{ \la1|p_2|\zeta]  \over m^{2S} [1\zeta] } 
\bigg [  \mls{3}\mrs{2}^{2S}  
+  \sum_{n=1}^{2S} \delta_n \left( \mls{3}\mrs{2} - \mla{3}\mra{2} \right)^n 
\mls{3}\mrs{2}^{2S-n}    \bigg ]
, \nnl 
    &\cM\left(1_\gamma^+ \bold2_X \bold3_{\bar X}\right)=  - \sqrt{2}  q_X  
{\la\zeta|p_2|1]  \over  m^{2S} \la1\zeta\ra   }
\bigg [ \mla{3}\mra{2}^{2S}  
+    \sum_{n=1}^{2S}\bar \delta_n \left(\mla{3}\mra{2} -\mls{3}\mrs{2}\right)^n
\mla{3}\mra{2}^{2S-n} 
\bigg ] 
, \nnl 
    &\cM\left(1_h^- \bold2_X \bold3_{\bar X}\right)=   
    -\frac{1}{\mpl} 
    \frac{\la1|p_2|\zeta]^2}{m^{2S}[1\zeta]^2 }\left[\mls{2}\mrs{3}^{2S}
    +\sum_{n=2}^{2S} \delta_n
    \left( \mls{3}\mrs{2}  - \mla{3}\mra{2}\right)^n
    \mls{3}\mrs{2}^{2S-n}  \right]
  , \nnl   
    &\cM\left(1_h^+ \bold2_X \bold3_{\bar X}\right)=      
    -\frac{1}{\mpl}\frac{\la\zeta|p_2|1]^2}{m^{2S}\la1\zeta\ra^2}
    \left[\mla{2}\mra{3}^{2S}
    +\sum_{n=2}^{2S}\bar \delta_n \left(\mla{3}\mra{2}-\mls{3}\mrs{2}\right)^n \mla{3}\mra{2}^{2S-n}\right] 
. \end{align}
where $q_X \equiv Q_X e$ is the electric charge, and $|\zeta \ra$ ($|\zeta ]$) is a reference spinor that is not parallel to $|1 \ra$ ($|1]$).
All momenta in \cref{eq:CA_M3} are treated as incoming. 
We omit the symmetrization over massive little group indices, along with the indices themselves. 
The first term in each square bracket stands for minimal coupling, while non-mininal couplings are parametrized by $\delta_n$: 
the (anomalous) dipole coupling is proportional to $\delta_1$, the quadrupole is proportional to $\delta_2$, etc.\footnote{%
Notice that, as shown in Ref.~\cite{Chung:2018kqs,Falkowski:2020aso}, the dipole term is  absent in gravity on general grounds.} 
We will be interested in deriving Compton amplitudes up to linear order in $\delta_n$.
The residues over the different kinematic channels are computed by gluing together two 3-particle amplitudes through the exchange of a massive particle (or eventually a graviton).
Given the residues, the Compton amplitudes can be reconstructed up to contact terms, which 
we fix by requiring the softest possible asymptotic behavior at $E \to \infty$. 

We start from the same-helicity Compton amplitude $\cM\left(1_{\gamma}^-,2_{\gamma}^-,3_X,4_{\bar X}\right)$, found via the $t$-channel residue
\begin{align}
    &\text{Res}_{t \rightarrow m^2}\cM\left(1_{\gamma}^- 2_{\gamma}^- \bold 3_X \bold 4_{\bar X}\right)
    = -
    \cM\left(1_\gamma^- \bold 3_X\rightarrow \bold t_X\right)
    \cM\left(2_\gamma^- \bold t_X \bold 4_{\bar X}\right)
    \notag\\
    &=\frac{(-1)^{2S}2 q_X^2 \la12\ra^2}{m^{4S-2}(u-m^2)}\sum_{k=0}^{2S} \delta_k(\mls{4}\mrs{t}-\mla{4}\mra{t})^k \mls{4}\mrs{t}^{2S-k} \sum_{n=0}^{2S}\delta_n(\mls{t}\mrs{3}+\mla{t}\mra{3})^n \mls{t}\mrs{3}^{2S-n},
, \end{align}
and the $u$-channel one obtained from the above by 
$t \leftrightarrow u$. 
In this context, $\delta_0 \equiv 1$.  
The convention for contracted little group indices of $|\mra{t}$ and $|\mrs{t}$ is that $2S$ leftmost carry the upper index, and the $2S$ rightmost carry the lower index. 
Using some spinor manipulations, 
in particular 
\begin{align}   
\label{eq: t-channel spinor contraction}
\mls{4}\mrs{t}  \mls{t}\mrs{3} = 
m \mls{4} \mrs{3}
, \quad 
\mla{4}\mra{t}  \mla{t}\mra{3} = 
-m \mla{4} \mra{3}
, \quad 
\mls{4}\mrs{t}  \mla{t}\mra{3}  = 
- \mls{4}| \bar t |\mra{3} 
, \quad  
\mla{4}\mra{t}  \mls{t}\mrs{3}  = 
 \mla{4}| t |\mrs{3} 
,  \end{align} 
and the identity\footnote{The second term vanishes in the $t$-channel, where this relation is used.}
\begin{align}
    \label{eq: t-channel spinor combination}
    \mla{3}| p_1 |\mrs{4}+\mla{4} |p_1| \mrs{3}=
    -\frac{1}{m}\mla{3}|p_1 p_2 |\mra{4}
    -\frac{t-m^2}{m}\mla{3}\mra{4},
 \end{align}
we reconstruct the same helicity Compton amplitudes
\begin{align}
\label{eq:Same sign Compton}
&\cM\left(1_{\gamma}^- 2_{\gamma}^- \bold 3_\phi \bold 4_{\bar \phi}\right)=
\frac{2 q_X^2 m^2 \la12\ra^2}{(t-m^2)(u-m^2)} 
,\nnl 
&\cM\left(1_{\gamma}^- 2_{\gamma}^- \bold 3_\psi \bold 4_{\bar \psi}\right)=
\frac{2 q_X^2 m  \la12\ra^2}{(t-m^2)(u-m^2)}
\bigg [\mls{3}\mrs{4} 
+ \delta_1 \big ( \mls{3}\mrs{4} - \mla{3}\mra{4} \big )    \bigg ] 
\notag\\
& +  \delta_1  {2 q_X^2 \la12\ra \over m}
\left[\frac{\la1\mra{3} \la2 \mra{4}}{t-m^2}-\frac{\la1 \mra{4}\la2 \mra{3}}{u-m^2}\right]
+   \mathcal{O}\left(\delta_1^2\right)  
,\nnl 
&\cM\left(1_{\gamma}^- 2_{\gamma}^- \bold3_V \bold4_{\bar V}\right)=
\frac{2 q_X^2 \la12\ra^2}{(t-m^2)(u-m^2)}
\bigg [ 
\mls{3}\mrs{4}^2 
+ \delta_1 \big ( \mls{3}\mrs{4} - \mla{3}\mra{4} \big ) \mls{3}\mrs{4} 
\notag \\
&
+ \delta_2 \big ( \mls{3}\mrs{4} - \mla{3}\mra{4} \big )^2 
-\frac{ \delta_2}{m^2}\big (
\mla{3}|p_1|\mrs{4}\mla{3}|p_2|\mrs{4}
+\mla{4}|p_1|\mrs{3}\mla{4}|p_2|\mrs{3}
\big )
\bigg ]
   \notag\\
&+\frac{2 q_X^{2} \la12\ra}{m^2 }
\bigg ( \delta_1 \mls{3}\mrs{4}
+ \delta_2 \big ( \mls{3}\mrs{4} - \mla{3}\mra{4}  \big)  \bigg ) 
\left(
\frac{\la1\mra{3}\la2\mra{4}}
{t-m^2}
-\frac{\la1\mra{4}\la2\mra{3}}
{u-m^2}
\right)
+\mathcal{O}\left(\delta_n^2\right)  
. \end{align}
The double plus Compton amplitudes can be obtained from the above by 
\begin{equation}
    \cM\left(1_x^+ 2_x^+ \bold3_X^{h_3} \bold4_{\bar X}^{h_4}\right)= 
    \left[\cM\left((-1)_x^- (-2)_x^- (-\bold4)_{ X}^{-h_4} (-\bold3)_{ \bar X}^{-h_3}\right)\right]^*,
\end{equation}
where $x=\gamma,h$, which amounts to exchanging square and angle brackets and conjugating complex parameters.

For the same-sign gravi-Compton amplitudes he principles are the same, with the additional complication due to a non-vanishing $s$-channel residue:
\begin{align}
&\text{Res}_{s \rightarrow 0}
\cM\left(1_{h}^- 2_{h}^- \bold3_X \bold4_{\bar X}\right) =
- \cM\left(1_h^- 2_h^-\rightarrow s_h^- \right)
\cM\left(s_h^- \bold3_X \bold4_{\bar X}\right)
\notag \\ 
&= { m^{4-2S} \langle 1 2 \rangle^4   
    \over  \mpl^2  (t-m^2)  (u-m^2)  }  
\left[\mls{3}\mrs{4}^{2S}+\sum_{n=2}^{2S} \delta_n \left(\mls{4}\mrs{3} - \mla{4}\mra{3}\right)^n\mls{4}\mrs{3}^{2S-n}\right]
   ,  \notag\\
    &\text{Res}_{t \rightarrow m^2}
    \cM\left(1_{h}^-2_{h}^-\bold 3_X\bold4_{\bar X}\right)=
    -\cM\left(1_h^- \bold3_X\rightarrow \bold t_X\right)\cM\left(2_h^- \bold t_X \bold 4_{\bar X}\right)=
    \notag\\
    &= { (-)^{2S}m^{4-4S} \langle 1 2 \rangle^4   
    \over  \mpl^2  s  (u-m^2)  }  
    \left[\mls{t}\mrs{4}^{2S}
    +\sum_{n=2}^{2S} \delta_n
    (\mls{4}\mrs{t}-\mla{4}\mra{t})^n\mls{4}\mrs{t}^{2S-n}\right]
    \left[\mls{3}\mrs{t}^{2S}+\sum_{k=2}^{2S} \delta_k(\mla{t}\mra{3}+\mls{t}\mrs{3})^k\mls{t}\mrs{3}^{2S-k}\right ] 
    , \end{align}
and the $u$-residue is obtained from the $t$ channel by $1 \leftrightarrow 2$.
We rewrite them making again use of 
\cref{eq: t-channel spinor contraction,eq: t-channel spinor combination} so as to reconstruct the double-minus gravi-Compton amplitude up to  linear quadrupole corrections:
\begin{align}
\label{eq:Same sign graviCompton}
    &\cM\left(1_h^- 2_h^- \bold 3_X \bold 4_{\bar X}\right)=
    \frac{m^{4-2S} \la12\ra^4}
    {\mpl^2 s (t-m^2)(u-m^2)}
\left(\mls{3}\mrs{4}^{2S}
+\delta_2 \big (\mls{4}\mrs{3}-\mla{4}\mra{3}\big )^2
\mls{4}\mrs{3}^{2S-2}\right)
\notag\\
&-\frac{\delta_2\la12\ra^2\mls{4}\mrs{3}^{2S-2}}{\mpl^2 m^{2S-2}}
\left(\frac{2\la1\mra{3}\la2\mra{4}\la2\mra{3}\la1\mra{4}}{ (t-m^2)(u-m^2)}+\frac{\la1\mra{3}^2\la2\mra{4}^2}{m^2 (t-m^2)}+\frac{\la1\mra{4}^2\la2\mra{3}^2}{m^2(u-m^2)}\right)+   \mathcal{O}\left(\delta_2^2\right), 
\end{align}
where $\delta_2$ is non-zero only if the spinning particle $X$ has $S\geq1$. 
Here, we present only the result involving  quadrupole corrections because in this paper we do not attempt to calculate corrections to Wilson coefficients from higher multipoles.
In~\cref{app:RESULT} we also report linear octupole corrections to the Compton amplitude.

We move to the opposite helicity Compton amplitude $M\left(1_{\gamma}^-,2_{\gamma}^+,3_X,4_{\bar X}\right)$. 
These require to compute the residues 
\begin{align}
    &\text{Res}_{t \rightarrow m^2}\cM\left(1_{\gamma}^- 2_{\gamma}^+ \bold3_X \bold4_{\bar X}\right)=
   -\cM\left(1_\gamma^- \bold3_X\rightarrow \bold t_X\right)\cM\left(2_\gamma^+ \bold t_X  \bold4_{\bar X}\right)
    \notag\\
    &= (-1)^{2S}\frac{2 q_X^2 \la 1|p_3|2]^2 }{m^{4S} (u-m^2)} 
    \sum_{k=0}^{2S}\bar \delta_k(\mla{4}\mra{t}-\mls{4}\mrs{t})^k \mla{4}\mra{t}^{2S-k} \sum_{n=0}^{2S}\delta_n(\mla{t}\mra{3}+\mls{t}\mrs{3})^n \mls{t}\mrs{3}^{2S-n}
    ,\notag\\
    &\text{Res}_{u \rightarrow m^2}\cM\left(1_{\gamma}^- 2_{\gamma}^+ \bold3_X \bold4_{\bar X}\right)= - \cM\left(2_\gamma^+ \bold3_X\rightarrow (-\bold u)_X\right)\cM\left(1_\gamma^- (-\bold u)_X  \bold4_{\bar X}\right)
    \notag\\
    &=
    (-1)^{2S}\frac{2 q_X^2 \la 1|p_3|2]^2 }{m^{4S} (t-m^2)} 
\sum_{k=0}^{2S}\delta_k(\mla{4}\mra{u}+\mls{4}\mrs{u})^k \mls{4}\mrs{u}^{2S-k} \sum_{n=0}^{2S}\bar \delta_n(\mla{u}\mra{3}-\mls{u}\mrs{3})^n \mla{u}\mra{3}^{2S-n}. 
    \end{align}
Using in particular
    \begin{align}
    &\la1|p_3|2]\mla{4}|p_t|\mrs{3}=
    -m^2\left(\la1\mra{3}[2\mrs{4}+\la1\mra{4}[2\mrs{3}\right)
    -(t-m^2) \la1\mra{4} [2\mrs{3}  ,
    \notag\\
&\la1|p_3|2]\mla{3}|p_u|\mrs{4}=
m^2\left(\la1\mra{3}[2\mrs{4}+\la1\mra{4}[2\mrs{3}\right)
    +(u-m^2) \la1\mra{3} [2\mrs{4}  
    ,\end{align}
and 
\begin{align}  
m \big ( \la1\mra{3}[2\mrs{4}+\la1\mra{4}[2\mrs{3} \big ) 
= & 
[12] \la1\mra{3} \la1\mra{4}
-   \la1|p_3|2]  \mla{4} \mra{3}
= - \la12\ra [2\mrs{3} [2\mrs{4}
- \la1|p_3|2]  \mls{4} \mrs{3}     
,  \end{align}  
the residues can be brought to a form that allows one to reconstruct the amplitudes for $S \leq 1$
    \begin{align}
        \label{eq:Opposite sign Compton}
    &\cM\left(1_\gamma^- 2_\gamma^+ \bold3_X \bold4_{\bar X}\right)= \frac{2 q_X^2  \la1|p_3|2]^{2-2S}}{(t-m^2)(u-m^2)}\left[
    \left(\la1\mra{3}[2\mrs{4}+\la1\mra{4}[2\mrs{3}\right)^{2S}
    \right.    \notag\\   &\left. 
+\sum_{n=1}^{2S}
\frac{\delta_n[12]^n\la1\mra{3}^n\la1\mra{4}^n
+ (-1)^n \bar \delta_n \la12\ra^n[2\mrs{3}^n [2\mrs{4}^n }
{m^n}
\left(
\la1\mra{3}[2\mrs{4}+\la1\mra{4}[2\mrs{3}\right)^{2S-n}
\right]
\nnl & 
+   \mathcal{O}\left(\delta_n^2\right) 
.    \end{align}

Finally, the opposite sign gravi-Compton amplitude is found through the residues
\begin{align}
    &\text{Res}_{s \rightarrow 0}\cM\left(1_{h}^- 2_{h}^+ \bold3_X \bold4_{\bar X}\right)=
    - \sum_{s'=\pm} \cM\left(1_h^- 2_h^+\rightarrow s_h^{s'}\right)\cM\left(s_h^{s'} \bold3_X \bold4_{\bar X}\right)
    \notag\\
    &=  {\langle 1| p_3 |2]^4      
    \over  \mpl^2 m^{2S} (t - m^2)  (u - m^2)   } 
    \left[\mla{3}\mra{4}^{2S}
    +\sum_{n=2}^{2S} 
     \bar \delta_n
    \left(\mla{4}\mra{3}-\mls{4}\mrs{3}\right)^n\mla{4}\mra{3}^{2S-n}\right]_{[12]\to 0}
  \nnl & =    
   {\langle 1| p_3 |2]^4      
    \over  \mpl^2 m^{2S} (t - m^2)  (u - m^2) }  
    \left[\mls{3}\mrs{4}^{2S}
    +\sum_{n=2}^{2S} 
      \delta_n
    \left(\mls{4}\mrs{3}-\mla{4}\mra{3}\right)^n\mls{4}\mrs{3}^{2S-n}\right]_{\langle 12 \rangle \to 0 }
    \notag\\
    &\text{Res}_{t \rightarrow m^2}\cM\left(1_{h}^-2_{h}^+\bold 3_X\bold4_{\bar X}\right)=
      -  \cM\left(1_h^- \bold3_X\rightarrow \bold t_X\right)\cM\left(2_h^+ \bold t_X \bold 4_{\bar X}\right)
      \notag\\
    &=
   {\langle 1| p_3 |2]^4      
    \over  \mpl^2 m^{4S} (t - m^2)  (u - m^2) }
\left[\mla{t}\mra{4}^{2S}+\sum_{n=2}^{2S}\bar \delta_n(\mla{4}\mra{t}-\mls{4}\mrs{t})^n\mla{4}\mra{t}^{2S-n}\right]\cdot\notag\\  &\cdot\left[\mls{3}\mrs{t}^{2S}+\sum_{k=2}^{2S} \delta_k(\mla{t}\mra{3}+\mls{t}\mrs{3})^k\mls{t}\mrs{3}^{2S-k}\right] 
,\end{align}
and the $u$ residue is obtained from the $t$ residue by $3 \leftrightarrow 4$. 
The same spinor manipulations used for the previous case allow us to reconstruct for $S \leq 2$
\begin{align}
   \label{eq:Opposite sign graviCompton}
   &\cM\left(1_h^- 2_h^+ \bold3_X \bold4_{\bar X}\right) = \frac{\la1|p_3|2]^{4-2S}}{\mpl^2 s (t-m^2)(u-m^2)}\left[\left(\la1\mra{3}[2\mrs{4}+\la1\mra{4}[2\mrs{3}\right)^{2S}+\right.\notag\\
    &\left.+(-1)^{2S}\sum_{n=2}^{2S}\frac{\delta_n([12]\la1\mra{3}\la1\mra{4})^n+(-1)^n\bar \delta_n(\la12\ra[2\mrs{3}[2\mrs{4}])^n}{m^n}\left(\la1\mra{3}[2\mrs{4}+\la1\mra{4}[2\mrs{3}\right)^{2S-n}\right] 
    \nnl & 
+   \mathcal{O}\left(\delta_n^2\right)
   . \end{align}

With the explicit form of the Compton amplitudes at hand, it is straightforward to extract the high-energy behaviour of the amplitudes by simple dimensional inspection of \cref{eq:Same sign Compton,eq:Same sign graviCompton,eq:Opposite sign Compton,eq:Opposite sign graviCompton}. 
Let as abbreviate 
$\cM_x^{\rm SS} = \cM[1_x^\pm,2_x^\pm,3_X,4_{\bar X}]$,
$\cM^{\rm OS} = \cM[1_x^-,2_x^+,3_X,4_{\bar X}]$ for $x=\gamma,h$. 
Consider generic kinematics with 
$s \sim t\sim u \sim E^2$ 
and take the limit $E \to \infty$. 
Recall that in the limit $\delta_n \to 0$ one has
$\cM_\gamma^{\rm SS} \sim 
(m/E)^{2-2S}$, 
$\cM_\gamma^{\rm OS} \sim E^0$, signaling that minimal interactions of photons with matter of spin 
$S \leq 1$ are renormalizable. 
In GR one finds
 $\cM_h^{\rm SS} \sim 
 E^{2S-2}/(\mpl^2 m^{2S-4})$, 
$\cM_h^{\rm OS} \sim (E/\mpl)^2$, 
the latter hitting strong coupling at $E \sim \mpl$, making GR an EFT with the cutoff $\mpl$. 
Now, in the presence of multipoles, the asymptotics change: 
\begin{align}
\cM_\gamma^{\rm SS} \sim & 
\delta_n {E^{2S} \over  m^{2S} }, 
\qquad
\cM_\gamma^{\rm OS} \sim  \delta_n {E^n \over m^n}, 
\qquad
S \leq 1, 
\quad 
1 \leq n \leq 2S
\nnl 
\cM_h^{\rm SS} \sim &
\delta_2 {E^{2S+2} \over \mpl^2 m^{2S}},
\qquad
\cM_h^{\rm OS} \sim  
\delta_2 {E^4 \over \mpl^2 m^2},
\qquad
S \leq 2
. \end{align}
Note that this behavior cannot be softened by adding contact terms to our Compton amplitudes, as those would induce $\sim E^{2S+2} (E^{2S+4})$ or harder asymptotic behavior in the electromagnetic (gravitional) setting.
It follows that the Compton amplitudes in the presence of multipoles hit strong coupling at the scale  $\Lambda_s \sim m/\delta_n^{1/2S}$. 
Identifying the strong coupling scale with the cutoff $\Lambda$, 
we can assign the power counting 
\begin{align}
\label{eq:COMPTON_powercounting}
\delta_n \sim {m^{2S} \over \Lambda^{2S}}
 \end{align}
to the multipole couplings in the electromagnetic theory, 
such that $\cM_\gamma^{\rm SS} \sim (E/\Lambda)^{2S}$. 
We will use the same power counting for quadrupole corrections in the gravitational theory, except that in this case we set $\Lambda = \mpl$ for simplicity. 
Then $\cM_h^{\rm SS} \sim (E/\mpl)^{2S+2}$ and the strong coupling scale is identified with the Planck scale. 
Note that, in the presence of multipoles,  the opposite-sign (gravi)Compton amplitudes have always the same or softer behavior at $E \to \infty$ compared to the same-sign ones.
The power counting in \cref{eq:COMPTON_powercounting} will play an important role in the following, allowing us to control the contributions from rational terms and tadpoles to the Wilson coefficients of EHEFT and GREFT. 

\section{Wilson coefficients in Euler-Heisenberg 
EFT} 
\label{sec:EH}
In this section we match the Wilson coefficients of EHEFT to the class of UV completion described in \cref{sec:COMPTON}. 
Using unitarity methods, we compute the one-loop amplitudes involving four photons with a massive spinning particle $X$ inside the loop. 
In particular, our method allows us to reconstruct the relevant part of the amplitude (that is, box, triangle, and bubble coefficients) with only two-particle cuts\footnote{%
In the massless context, similar approach was introduced in Refs.~\cite{Britto:2005ha,Britto:2006sj}.} 
in $d=4$, as opposed to the more familiar recursion requiring triple and quadruple cuts as well. 
Effectively, cuts in just one or two channels need to be calculated (depending on the helicity configuration), the remaining being trivially obtained by crossing. 

We first decompose the loop amplitude in a basis of scalar integrals:
\begin{equation}
\label{eq:1-loop decomposition}
    \cM_{\text{1-loop}}=\sum_i c_{\Box}^iI_{\Box}^i+\sum_i c_{\triangleright}^iI_\triangleright^i+\sum_2 c_{\circ}^iI_{\circ}^i+c \Delta I+R,
\end{equation}
where $I_\Box^i$, $I_\triangleright^i$, $I_\circ^i$ are respectively box, triangle, and bubble scalar integrals defined in \cref{APP:scalar integrals}, and $\Delta I$ is an eventual combination of the tadpole and massless bubble contributions.
We then compute the discontinuity over the normal threshold of a Mandelstram invariant of both sides of the decomposition: 
\begin{equation}
\label{eq:1-loop decomposition disc}
    \text{Disc}^s\cM_{\text{1-loop}}=\sum_i c_{\Box}^i\text{Disc}^s I_{\Box}^i+\sum_i c_{\triangleright}^i\text{Disc}^s I_\triangleright^i + \sum_i c_{\circ}^i\text{Disc}^s I_{\circ}^i.
\end{equation} 
The discontinuities of the scalar integrals on the right-hand side are listed in \cref{app:INT}, while the discontinuity of the original loop amplitude is found by integrating the product of two Compton amplitudes over the two-particle phase space. 
Matching independent non-analytic structures on either sides of \cref{eq:1-loop decomposition disc}, we are able to reconstruct all box, triangle, and  bubble coefficients. 
The rational terms and tadpoles can be fixed by cancelation of non-physical poles and power-counting arguments. 
Finally, taking the $m\rightarrow\infty$ limit and keeping the leading orders in $\frac{1}{m}$, we match the amplitudes in the UV completion to the four-photon amplitudes in EHEFT. 
The latter, starting from \cref{eq:INT_Leheft}, can be written as
\begin{align}
    &\cM_-= \cM\left(1_\gamma^-,2_\gamma^-,3_\gamma^-,4_\gamma^-\right)=
    C_-\left(\la12\ra^2\la34\ra^2+\la13\ra^2\la24\ra^2+\la14\ra^2\la23\ra^2\right),
    \label{eq:4-photons -- EH}
    \\
&\cM_+=\cM\left(1_\gamma^-,2_\gamma^-,3_\gamma^+,4_\gamma^+\right)=
    C_+\la12\ra^2[34]^2,
    \label{eq:4-photons -+ EH}
    \\
&\cM\left(1_\gamma^+,2_\gamma^+,3_\gamma^+,4_\gamma^+\right)=
C_-\left([12]^2[34]^2+[13]^2[24]^2+[14]^2[23]^2\right),
    \label{eq:4-photons ++ EH}
\end{align}
where $C_\pm=\frac{C_1\pm C_2}{2}$.
In the following we will illustrate in some detail how the method works using the well-known example of integrating out the massive scalar~\cite{Weisskopf:1936hya}. 
We then report the results for the matching up to spin $S=1$ with only minimal interactions (which have also been long known~\cite{Euler:1935qgl,Vanyashin:1965ple}). 
Finally, we discuss how the method can be adapted to the case of integrating out particles with $S\leq 1$ and non-minimal electromagnetic interactions.

\subsection{Minimal coupling}
\label{sec:EH_minimal}

We first consider the UV completion of EHEFT to be a scalar particle charged under electromagnetic interactions.  
Let us start by computing the double cut of the one-loop four-photon amplitude. 
Beginning with the all-minus configuration $\cM_-$,
we have
\begin{align}
   & \text{Disc}^s\cM_- = 
   i\int \text{d}\Pi_{XY}\cM\left(1_\gamma^- 2_\gamma^- (-\bold Y)_\phi (-\bold X)_{\bar\phi}\right)\cM\left(3_\gamma^- 4_\gamma^- \bold X_\phi \bold Y_{\bar\phi}\right)
   \notag\\
    &= 4ie^4m^4\la12\ra^2\la34\ra^2\int
    \frac{\text{d}\Pi_{XY}}
{ (2p_1 p_X)(2p_2 p_X) (2p_3 p_X)  (2p_4 p_X)  } 
,\end{align}
where $\text{d}\Pi_{XY}$ is the phase space element of the $XY$ two-particle state.  
Then, using the parametrization for the internal momenta defined in \cref{eq: parametrization s channel}, 
we find
\begin{equation}
    \text{Disc}^s\cM=
    -\frac{i  q_X^4 m^4 \la12\ra^2\la34\ra^2}{2\pi s^2 \la23\ra^2 [13]^2}\int\frac{\text{d}\alpha}{\alpha(1-\alpha)}
    \frac{1}{\alpha(1-\alpha)-\frac{1}{4y_s}}
    \frac{\text{d}z}{2\pi i}\frac{z}{(z-z_{s_-})(z-z_{s_+})(z-\tilde z_{s_-})(z-\tilde z_{s_+})},
\end{equation}
where $y_s\equiv\frac{s}{4m^2}$. 
The integral over $z$ encompasses a unit circle centered around the origin, and
\begin{align}
\label{eq: zplusminus s channel}
    &z_{s_\pm}\equiv\frac{u}{\la23\ra[13]}\frac{\alpha+x_s(1-\alpha)\pm\sqrt{[\alpha-x_s(1-\alpha)]^2+\frac{x_s}{y_s}}}{2\sqrt{\alpha(1-\alpha)-\frac{1}{4y_s}}},\\
    &\tilde z_{s_\pm}\equiv-\frac{u}{\la23\ra[13]}\frac{x_s\alpha+(1-\alpha)\pm\sqrt{[x_s\alpha-(1-\alpha)]^2+\frac{x_s}{y_s}}}{2\sqrt{\alpha(1-\alpha)-\frac{1}{4y_s}}},\\
\end{align}
with $x_s\equiv\frac{t}{u}$. One can show that $z_{s_-}$ and $\tilde z_{s_-}$ are always inside the $|z|<1$ integration contour, whereas $z_{s_+}$ and $\tilde z_{s_+}$ are always outside. Integrating over $z$ picks the two former residues, which yields, after some manipulation,
\begin{equation}
     \text{Disc}^s\cM_-=-\frac{i q_X^4  m^4 \la12\ra^2\la34\ra^2}{\pi s^3 u}\int_{\alpha_-}^{\alpha_+}\text{d}\alpha\left(\frac{1}{\alpha}+\frac{1}{1+\alpha}\right)\frac{1}{\sqrt{[\alpha-x(1-\alpha)]^2+\frac{x}{y}}}.
\end{equation}
Now, matching this to the dictionary in \cref{APP: dictionary}, we find
\begin{equation}
\label{eq:disc of M in EH with scalar}
     \text{Disc}^s\cM_-=\frac{8 q_X^4  m^4\la12\ra^2\la34\ra^2}{s^2}\left( \text{Disc}^s I_\Box^{st}+ \text{Disc}^s I_\Box^{su}\right),
\end{equation}
from which we reconstruct the full amplitude up to a rational term
\begin{equation}
\label{eq:M in EH with scalar}
    \cM_-=\frac{8 q_X^4 
 m^4\la12\ra^2\la34\ra^2}{s^2}\left(I_\Box^{st}+I_\Box^{su}+I_\Box^{tu}\right)+R.
\end{equation}
\cref{eq:M in EH with scalar} is the only solution that is consistent with \cref{eq:disc of M in EH with scalar} and its analogous versions in other kinematic channels, straightforwardly found by crossing\footnote{This is easily seen noticing that $\frac{\la12\ra^2\la34\ra^2}{s^2}=\frac{\la13\ra^2\la24\ra^2}{t^2}=\frac{\la14\ra^2\la23\ra^2}{u^2}$.}. 
Moreover, it is symmetric under the exchange of any two photons, by construction.
Tadpoles cannot appear in \cref{eq:M in EH with scalar} because they would introduce divergences to the amplitude, which cannot appear in a renormalizable theory as that would imply introducing $\sim F^4$ counterterms to the Lagrangian.

Expanding the scalar integrals in $1\over m$ we find
\begin{equation}
    \cM_-=\frac{q_X^4 \la12\ra^2\la34\ra^2}{4\pi^2 s^2} +R  + \dots 
, \end{equation}
where the dots have no kinematic singularities. 
To have a consistent amplitude, the rational term must cancel the $1\over s^2$ pole: 
\begin{equation}
\label{eq:EH_R0-}
R = -\frac{q_X^4\la12\ra^2\la34\ra^2}{4\pi^2 s^2} 
. \end{equation}
Now, the crucial point of this construction is that \cref{eq:EH_R0-} cannot contain any other pieces. 
In particular, it cannot contain any terms contributing to $C_-$. 
Indeed, by dimensional analysis, such terms would have to be of the order  
$\Delta R \sim  { \la12\ra^2\la34\ra^2 \over m^4} $, but that would make $\cM_-$ hit the unitarity limit  around $E \sim m$, contradicting the renormalizability of the theory.  

With the rational term fixed as in \cref{eq:EH_R0-}, expanding \cref{eq:M in EH with scalar} yields
\begin{equation}
    \cM_-=\frac{q_X^4}{480\pi^2 m^4}\left(\la12\ra^2\la34\ra^2+\la13\ra^2\la24\ra^2+\la14\ra^2\la23\ra^2\right)+\mathcal{O}\left(m^{-6}\right).
\end{equation}
We can then match $C_-$ in \cref{eq:4-photons -- EH} as
\begin{equation}
\label{C- spin0 EH}
    C_-=\frac{q_X^4}{480\pi^2 m^4}=\frac{\alpha^2}{30m^4}
, \end{equation}
where $\alpha \equiv {q_X^2 \over 4 \pi}$. 

The matching of the minimally helicity-violating (MHV) amplitude  $\cM_+$is similar, 
but we will need to compute the discontinuity of the loop amplitude over two different kinematic channels in order to fully reconstruct it.
The s-channel discontinuity is found exactly as before, yielding
\begin{equation}
\label{eq:DiscS spin0 in EH}
    \text{Disc}^s\cM_+=
    \frac{8 q_X^4 m^4\la12\ra^2[34]^2}{s^2}\left(\text{Disc}^s I_\Box^{st}+\text{Disc}^s I_\Box^{su}\right).
\end{equation}
We now need to compute the t-channel discontinuity as well, as it  cannot obtained by crossing the former in this case.
\begin{align}
   & \text{Disc}^t\cM_+=
   i\int \text{d}\Pi_{XY}\cM\left(1_\gamma^- 3_\gamma^+ (-\bold Y)_\phi (-\bold X)_{\bar\phi}\right)\cM\left(2_\gamma^- 4_\gamma^+ \bold X_\phi \bold Y_{\bar\phi}\right)=\notag\\
    &=i \la12\ra^2[34]^2
    \int\frac{\text{d}\Pi_{XY}X^{(0)}}
{  (2p_1 p_X)(2p_2 p_X) (2p_3 p_X)  (2p_4 p_X)} 
    \label{Eq:DiscT spin 0 EH}
\end{align}
with
\begin{equation}
\label{eq: X0 EH}
    X^{(0)} \equiv 
    4 q_X^4 \frac{\la1| p_X| 3]^2\la2| p_X |4]^2}{\la12\ra^2[34]^2}
.\end{equation}
Parametrizing the internal momenta analogusly as in~\cref{eq: parametrization s channel} but crossing 
$(2 \leftrightarrow 3)$, we can rewrite \cref{Eq:DiscT spin 0 EH} as
\begin{align}
\label{Eq:DiscT spin 0 EH final form}
    &X^{(0)}(z)=t^2\left(\alpha(1-\alpha)-\frac{m^2}{t}\right)\left(z(2\alpha-1)\frac{[14]}{[34]}+\sqrt{\alpha(1-\alpha)-\frac{m^2}{t}}\left[z^2\frac{\la23\ra[14]}{\la12\ra[34]}-1\right]\right)^2,\notag\\
    & \text{Disc}^t\cM_+=-\frac{i e^4\la12\ra^2[34]^2}{\pi t^3 u}\int\frac{\text{d}\alpha}{\alpha(1-\alpha)}\frac{X^{(0)}(z_{t_-})}{\sqrt{[\alpha-x_t(1-\alpha)^2]+\frac{x_t}{y_t}}}
\end{align}
where\footnote{In order to find \cref{Eq:DiscT spin 0 EH final form} we again used that only two of the $z$ poles are inside the integration contour, along with the property $X^{(0)}(z_{t_-},\alpha)=X^{(0)}(\tilde z_{t_-},1-\alpha)$}
$z_{t-}$ is defined from \cref{eq: zplusminus s channel} by crossing.
Now it is straightforward to match this discontinuity to the ones of the basis of scalar integrals using the dictionary in \cref{APP: dictionary}, leading to
\begin{align}
\text{Disc}^t\cM_+=&
q_X^4\la12\ra^2[34]^2\left[\frac{8m^4}{s^2}(\text{Disc}^t I^{st}_\Box+\text{Disc}^t I^{tu}_\Box)+\frac{4u^2t^2+16m^2 stu}{s^4}\text{Disc}^t I^{tu}_\Box-\right.\notag\\
     &\left.-t\frac{8tu+16m^2 s}{s^4}\text{Disc}^t I^{t}_\triangleright-4\frac{t-u}{s^3}\text{Disc}^t I^{t}_\circ\right].
\end{align}
The discontinuity in the u-channel is found by crossing this result, and together with \cref{eq:DiscS spin0 in EH} they allow us to reconstruct the full amplitude up to rational terms as
\begin{align}
    \cM_+=&e^4\la12\ra^2[34]^2\left[\frac{8m^4}{s^2}( I^{st}_\Box+ I^{tu}_\Box+I^{su}_\Box)+\frac{4u^2t^2+16m^2 stu}{s^4} I^{tu}_\Box-\right.\notag\\
     &\left.-\frac{8tu+16m^2 s}{s^4}\left( t I^{t}_\triangleright+ u I^{u}_\triangleright\right)-4\frac{t-u}{s^3} (I^t_\circ-I^{u}_\circ)\right]+R.
\end{align}
Notice that the form of the discontinuities in different channels provides valuable crosschecks for our computations, as already noted in \cite{Bern:2021ppb}, since both $s$- and $t$-channel cuts must yield the same $st$ box coefficient.

The rational term $R$ is fixed to cancel the non-physical poles produced by the scalar integrals.
The same argument as before excludes any contributions of $R$ to the Wilson coefficients. 
Fixing $R$, expanding the scalar integrals, and matching the result to \cref{eq:4-photons -+ EH} to identify the Wilson coefficient $C_+$ , one finds
\begin{equation}
    C_+=\frac{2\alpha^2}{45 m^4}.
\end{equation}
Combining this result with \cref{C- spin0 EH} we recover the result of Refs.~\cite{Weisskopf:1936hya,Preucil:2017wen}:
\begin{equation}
    C^{\rm SQED}_1=\frac{7\alpha^2}{90m^4},\qquad C^{\rm SQED}_2=\frac{\alpha^2}{90m^4}.
\end{equation}

\vspace{1cm}

The computation for spinning particles with $S=1/2,1$ is essentially equivalent to the scalar case. 
For the matching of $\cM_-$,
the only difference with the scalar case is an
 extra factor of 
\begin{equation}
\label{eq: contraction massive spinors angle-angle}
   \frac{\mla{X}\mra{Y}^{I_1 J_1}\mla{Y}\mra{X}_{J_1 I_1}\cdot... \cdot   \mla{X}\mra{Y}^{I_{2S} J_{2S}}\mla{Y}\mra{X}_{J_{2S} I_{2S}}}{m^{4S}}=
   (-)^{2S}(2S+1)
   , \end{equation}
where the little group indices $I_k$ and $J_k$ are implicitly fully symmetrized.
While matching $\cM_+$, there are only two changing factors. For the s-channel discontinuity, we have to take into account the overall contraction
\begin{equation}
      \frac{\mls{X}\mrs{Y}^{I_1 J_1}\mla{Y}\mra{X}_{J_1 I_1}\cdot... \cdot   \mls{X}\mrs{Y}^{I_{2S} J_{2S}}\mla{Y}\mra{X}_{J_{2S} I_{2S}}}{m^{4S}}.
\end{equation}
For the t-channel discontinuity, the only changing structure is the function $X^{(S)}(z)$ arising from the different Compton amplitudes, and the consequent matching with the discontinuities with the basis of scalar integrals. 
Here we just quote the final results for these amplitudes, listing the $X^{(S)}$ functions in
\cref{eq:RESULTS_XfunctionsPhotons}. 
For $S=\frac{1}{2}$, we find
\begin{align}
   & \cM_-=-\frac{16 q_X^4
   m^4\la12\ra^2\la34\ra^2}{s^2}\left(I_\Box^{st}+I_\Box^{su}+I_\Box^{tu}-\frac{1}{32\pi^2m^4}\right)+R
   ,\notag \\
   & \cM_+=
   -q_X^4 \la12\ra^2[34]^2\left[\frac{16m^4-8m^2 s}{s^2}( I^{st}_\Box+ I^{tu}_\Box+I^{su}_\Box)+\frac{32m^2 stu-4tu(t^2+u^2)}{s^4} I^{tu}_\Box-\right.\notag\\
     &\left.\qquad-\frac{32m^2 s-8(t^2+u^2)}{s^4}\left( t I^{t}_\triangleright+ u I^{u}_\triangleright\right)-8\frac{t-u}{s^3}\left(I^{t}_\circ-I^{u}_\circ\right)\right]+R,
\end{align}
leading, upon matching, to
\begin{equation}
        C^{\rm QED}_1=
        \frac{8\alpha^2}{45m^4}
        ,\qquad 
        C^{\rm QED}_2=
        \frac{14\alpha^2}{45m^4},
\end{equation}
in agreement with Refs.~\cite{Euler:1935qgl,Preucil:2017wen}. 
For $S=1$ we find
\begin{align}
   & \cM_-=
   \frac{24 q_X^4 m^4\la12\ra^2\la34\ra^2}{s^2}\left(I_\Box^{st}+I_\Box^{su}+I_\Box^{tu}-\frac{1}{32\pi^2m^4}\right)+R
   ,\notag \\
   & \cM_+=
   q_X^4 \la12\ra^2[34]^2\left[8\frac{3m^4-4m^2 s+ s^2}{s^2}( I^{st}_\Box+ I^{tu}_\Box+I^{su}_\Box)+\frac{48m^2 stu-4tu(4t^2+4u^2+5tu)}{s^4} I^{tu}_\Box+\right.\notag\\
     &
     \left.\qquad+8\frac{4(t^2+u^2)+5tu-6m^2 s}{s^4}\left( t I^{t}_\triangleright+ u I^{u}_\triangleright\right)-12\frac{t-u}{s^3}\left(I^{t}_\circ-I^{u}_\circ\right)\right]+R,
\end{align}
leading to
\begin{equation}
C^{\rm VQED}_1=\frac{29\alpha^2}{10m^4},\qquad C^{\rm VQED}_2=\frac{27\alpha^2}{10m^4}
, \end{equation}
in agreement with Refs.~\cite{Vanyashin:1965ple,Preucil:2017wen}.

\subsection{Multipoles}
\label{sec:EH_multipoles}

In the presence of multipole (dipole and quadrupole) couplings of matter to photons, the computation of EHEFT Wilson coefficients roughly parallels the minimal coupling case. 
There is one important difference though. 
Since the UV theory is now itself an EFT, we cannot use the renormalizability argument to harness the rational term and tadpole contributions to the 4-photon amplitude. 
Nevertheless, the argument can be adapted here using the EFT power counting, due to the fact that EFTs are renormalizable order by order in  $1/\Lambda$ expansion. 

Consider a rational term in the 4-photon amplitude contributing to a Wilson coefficient of dimension $[C]=[\rm mass]^{-n}$ and proportional to the multipole moment $\delta$ in some integer power $k$. 
By dimensional analysis it has to scale as 
$C \sim \delta^k/m^n$.
At large energies it will behave as
$R \sim \delta^k E^n/m^n$ and hit strong coupling at the scale 
$\Lambda_s \sim m/\delta^{k/n}$. 
Given the power counting in \cref{eq:COMPTON_powercounting}, this is 
 $\Lambda_s \sim \Lambda (m/\Lambda)^{1- 2Sk/n}$.  
If $2 S k \geq n$, such a rational term reaches the unitarity limit at the scale $\Lambda$ or above, and therefore it is perfectly allowed. 
Conversely,  for 
 \begin{align}
 \label{eq:EH_calculability}
    2 S k <   n 
,  \end{align}
such a rational term cannot appear in an EFT with a cutoff $\Lambda$. 
The same argument works for the tadpole terms.
Multipole corrections satisfying \cref{eq:EH_calculability} are {\em calculable} with our method, in the sense that they can be calculated from the knowledge of the cut-constructible part of the amplitude (boxes, triangles, and bubbles). 
This argument is equivalent to requiring  that the amplitudes in our EFT have a healthy $m \to 0$ limit.

Focusing on the Wilson coefficients  $C_{1,2}$ in \cref{eq:EH_calculability}, corresponding to $n=4$, we can see that for $S=1/2$, 
effects up to the cubic order in $\delta_1$ are calculable. 
For spin-1 particles, on the other hand,  only linear effects in $\delta_1$ and $\delta_2$ are calculable. 
This may appear as a serious restriction of the method, but from the EFT point of view it is not. 
The point is that $\cO(\delta_1^4)$ contributions for $S=1/2$, or $\cO(\delta_i^2)$ ones for $S=1$ are of the same order as the ${\cal L} \supset C_\Lambda F^4$ operators, which should be anyway included in the theory with an unknown coefficient $C_\Lambda  \sim \Lambda^{-4}$ according to the EFT power counting. 
Indeed, in general  we expect $C_\Lambda$ to be generated by the same physics that generates the multipole couplings $\delta_i$. 
In summary, our method only captures IR contributions to the Wilson coefficients that are parameterically enhanced by powers of $\Lambda/m$ compared to those coming from unknown physics above $\Lambda$. 
Note that, according to \cref{eq:EH_calculability}, calculability improves as we go to higher and higher orders in Wilson coefficients. 

Given that discussion, the rest of the calculations proceeds along the lines sketched in the previous subsection.
Integrating out a Dirac fermion interacting non-minimally with the photon 
we obtain the Wilson coefficients
\begin{align}
\label{eq:EH_C12-fermion}
    &C_1^{\left(1/2\right)}=
     \frac{q_X^4}{90\pi^2 m^4}
     +\mathcal{O}\left(\delta_1^4\right),\notag\\
    &C_2^{(1/2)}=
    \frac{q_X^4} {360\pi^2m^4}
    \bigg ( 7  + 60 \delta_1 + 150  \delta_1^2  + 120  \delta_1^3 \bigg ) 
    +\mathcal{O}\left(\delta_1^4\right)
\end{align}
Integrating out a massive vector with general dipole and quadrupole electromagnetic interactions
we obtain
\begin{align}
\label{eq:EH_C12-vector}
    &C_1^{\left(1\right)}=
     \frac{q_X^4}{\pi^2 m^4} \bigg  ( 
\frac{29}{160}+\frac{11 \delta_1}{12}
+\frac{21 \delta_2}{12} \bigg )  +\mathcal{O}\left(\delta_i^2\right)
  ,\notag\\
    &C_2^{(1)}=
     \frac{q_X^4}{\pi^2 m^4} \bigg  ( 
     \frac{27}{160}
     +\frac{2 \delta_1 }{3}
     \bigg ) 
     +\mathcal{O}\left(\delta_i^2\right) 
. \end{align}

Since our method includes construction of complete one-loop 4-photon amplitudes in the matter theory, it is straightforward to also extract  Wilson coefficients of higher-dimensional operators in EHEFT~\cite{Karplus:1950zz,Costantini:1971cj,Yang:1994nu,Henriksson:2021ymi}. 
The results for selected operators up to dimension~12 are summarized in \cref{tab:higher order coefficients EH} in \cref{app:higherorder}.

\subsection{Higher spins}

When integrating out particles with higher spins, there is a conceptual barrier that prevents our computation to work. 
Namely, our results for the Compton amplitudes found in \cref{sec:COMPTON}, even for non-minimal interactions, are incomplete for $S>1$. 
In fact, for higher spins, a non-physical pole at $\la1|p_3|2] \to 0$ appears, and we need extra terms to cancel it~\cite{Falkowski:2020aso,Chiodaroli:2021eug,Aoude:2022trd}. 
The choice of how to deal with the spurious pole is not unique (due to the usual ambiguity of contact terms), but it always leads to the Compton amplitudes growing at least as 
$\cM \sim q_X^2 (E/m)^{4S-2}$ at high energy~\cite{Porrati:2008ha,Falkowski:2020aso}. 
This signals that the higher-spin theory is always an EFT with cutoff  $\Lambda \sim m/|q_X|^{1\over 2S-1}$.
For a small enough $q_X$ it is possible to arrange for $\Lambda \gg m$, leading to a non-trivial validity range for the EFT.  
However, in this setting, nothing prevents the tadpole and rational terms to display a 
$\frac{q_X^4}{m^n}$ behaviour for some $n$, and eventually contribute to the Wilson coefficients at the same order as the cut-constructible terms.
Since the former are not calculable with our method, the Wilson coefficients cannot be determined. 
At the same time, one expects the contributions to the Wilson coefficients scaling with powers of $1/m$ from the unknown UV completion of the higher-spin theory above $\Lambda$, which would be of the same order or larger as those from integrating out the higher-spin particle. 

We remark however that there exists a compact expression for the higher-spin contribution with arbitrary $S$ to the Wilson coefficient $C_-$. 
That calculation depends only on the same-sign Compton amplitude, which for minimal couplings takes the simple form 
\begin{align}
\cM\left(1_{\gamma}^- 2_{\gamma}^- \bold 3_X \bold 4_{\bar X}\right)= & 
\frac{2 q_X^2 m^2 \la12\ra^2}{(t-m^2)(u-m^2)}  {[34]^{2S} \over m^{2S}} + {\rm contact} \, \, {\rm terms}
, \end{align}
and does not exhibit any spurious poles for arbitrary integer or half-integer $S \geq 0$. 
Assuming the absence rational and tadpole contributions and ignoring the contact terms one finds 
\begin{equation}
 C_-= 
(-)^{2S} (2S+1)  \frac{\alpha^2}{30 m^4}
.\end{equation}
This just counts the number of degrees of freedom of the higher-spin particle.
On the other hand, generic contributions to $C_+$ scale with higher powers of $S$. 
One then expects, given our assumptions, that the higher-spin contributions to the leading EHEFT Wilson coefficients approach the diagonal $C_1 \approx C_2$.

\section{Wilson coefficients in General Relativity EFT}  
\label{sec:GREFT}

Matching between the theory with a matter particle coupled to gravity and the GREFT is conceptually equivalent to the one carried out in the previous section. 
In particular, the discussion of calculability in \cref{sec:EH_multipoles} is unchanged (with $\Lambda \to \mpl$ and  the power counting 
$\delta_2 \sim (m/\mpl)^{2S}$), and the condition for calculability of the quadrupole corrections remains the one in \cref{eq:EH_calculability}.
The most important technical differences are
\begin{itemize}
    \item Dipole corrections are absent in gravity, $\delta_1 = 0$.
    \item We are free consider matter particles up to spin two, because the corresponding Compton amplitudes are free of spurious poles.  
    \item There exists the cubic operator $R^3$ in GREFT that can be matched to the matter EFT. 
\end{itemize}
Concerning the last point, 
on-shell methods do
not allow to match three-point amplitudes, but the $R^3$ operator contributes to 4-point amplitudes, and can thus be detected from their matching.
One finds that the four-point graviton amplitudes, including contributions from $R^3$ and $R^4$ operators, are given by~\cite{vanNieuwenhuizen:1976vb,Bern:2017puu}
\begin{align}
 &\label{eq:4-gravitons --}
\cM[1_h^-,2_h^-,3_h^-,4_h^-] \equiv  \cM_- 
=   \frac{C_{4,1}-C_{4,2}}{\mpl^4}\left(\la12\ra^4\la34\ra^4+\la13\ra^4\la24\ra^4+\la14\ra^4\la23\ra^4\right)\notag\\
    &\qquad\qquad\qquad\qquad
    +10\frac{ C_3}{\mpl^4}\frac{\la12\ra\la34\ra}{s}\la12\ra\la34\ra\la13\ra\la24\ra\la14\ra\la23\ra\\
    &\label{eq:4-gravitons -+}
\cM[1_h^-,2_h^-,3_h^+,4_h^+] \equiv  \cM_+ =
\la12\ra^4[34]^4\left[\frac{1}{\mpl^2 stu}+\frac{C_{4,1}+C_{4,2}}{\mpl^4}\right]
\\
    &\label{eq:4-gravitons ---+}
    \cM[1_h^-,2_h^-,3_h^-,4_h^+]=
    -\frac{C_3}{\mpl^4}\la12\ra^2\la23\ra^2\la31\ra^2\frac{[4|p_1  p_2|4]^2}{stu}.
\end{align}
Notice that the last amplitude is not needed for matching of the $C_3$ Wilson coefficient, which can be found from the all-minus one, but it provides a useful consistency check.

Let us begin with minimal coupling  and start from the all-minus amplitude, which we can reconstruct in one go for all spins $S\leq2$. 
We only need to compute the discontinuity in one kinematic channel, as the others are obtained by crossing. 
Iterating our methods to compute the discontinuities and reconstructing the amplitude yield
\begin{equation}
    \cM_-=(-1)^{2S}(2S+1)\frac{2m^8}{\mpl^4 s^4}\left(I_\Box^{st}+I_\Box^{su}+I_\Box^{tu}\right)+R 
    .\end{equation}
Cancelling spurious poles
using the rational term, and expanding amplitude in powers of $1/m$,  we are left with
\begin{align}
    \cM_-=&\frac{(-1)^{2S}(2S+1)}{16\pi^2 \mpl^4}\left[-\frac{\la12\ra^2\la34\ra^2\la13\ra\la24\ra\la14\ra\la23\ra}{252s m^2}+\right.\notag\\
    &\left.+\frac{\la12\ra^4\la34\ra^4+\la13\ra^4\la24\ra^2+\la14\ra^4\la23\ra^4}{3780 m^4}\right]+\mathcal{O}\left(m^{-6}\right).
\end{align}
Matching to \cref{eq:4-gravitons --}
\begin{align}
    C_3=-\frac{(-1)^{2S}}{16\pi^2}\frac{(2S+1)}{2520m^2},\\
    C_{4,1}-C_{4,2}=\frac{(-1)^{2S}}{16\pi^2}\frac{(2S+1)}{3780m^4}.
\end{align}
These results agree with Refs.~\cite{Goon:2016mil,Bern:2021ppb}.  Again, the results make sense for arbitrary $S>2$, with the understanding that then there may be additional contributions from tadpoles and rational terms.
The matching to the amplitude in~\cref{eq:4-gravitons ---+}, which is analogous and we omit, yields a consistent result for $C_3$.

Moving onto the matching of \cref{eq:4-gravitons -+}, we now need to compute the discontinuities over the s- and t-channel separately. 
The t-channel discontinuities have all the form
\begin{equation}
 \text{Disc}_t \cM_+ =
    i \la12\ra^4[34]^4    
    \int\text{d}\Pi_{XY}\frac{X^{(S)}(z,\alpha)}
    {(2p_1 p_X)(2p_2 p_X) (2p_3 p_X) (2p_4 p_X)},
\end{equation}
where the $X^{(S)}(z,\alpha)$ functions are given in \cref{eq:RESULTS_XfunctionsGravity},
and they lead to the results 
\begin{align}
   & C_+^{(0)}=\frac{1}{16\pi^2}\frac{1}{3150m^4}
   ,\notag\\
   & C_+^{(\frac{1}{2})}=\frac{1}{16\pi^2}\frac{29}{25200m^4}
   ,\notag\\
   & C_+^{(1)}=\frac{1}{16\pi^2}\frac{31}{6300m^4}
   ,\notag\\
   & C_+^{(\frac{3}{2})}=\frac{1}{16\pi^2}\frac{1257}{37800 m^4}
   ,\notag\\
   & C_+^{(2)}=\frac{1}{16\pi^2}\frac{671}{1260 m^4}
.\end{align}
From these we compute the coefficients $C_{4,1}$, $C_{4,2}$, which can be found in \cref{tab:GREFT_C34}.

\renewcommand{\arraystretch}{1.25}
\begin{table}[tb]
    \centering
    \begin{tabular}{|c|c|c|c|}
        \hline
        $S$ & $C_3$ & $C_{4,1}$ & $C_{4,2}$ \\ \hline
        $0$ & $-\frac{1}{16\pi^2}\frac{1}{2520 m^2}$ & $\frac{1}{16\pi^2}\frac{11}{37800 m^4}$ & $\frac{1}{16\pi^2}\frac{1}{37800 m^4}$ \\ \hline
        $\frac{1}{2}$ & $\frac{1}{16\pi^2}\frac{1}{1260 m^2}$ & $\frac{1}{16\pi^2}\frac{47}{151200 m^4}$ & $\frac{1}{16\pi^2}\frac{127}{151200 m^4}$ \\ \hline
        $1$ & $-\frac{1}{16\pi^2}\frac{1}{1260 m^2}(1\textcolor{Green}{+231\delta_2}) $& $\frac{1}{16\pi^2}\frac{1}{4200 m^4}(12+245 \delta_2)$ & $\frac{1}{16\pi^2}\frac{1}{12600 m^4}(26-175 \delta_2)$ \\ \hline
        $\frac{3}{2}$ & $\frac{1}{16\pi^2}\frac{1}{630 m^2}(1\textcolor{Green}{-420 \delta_2})$ & $\frac{1}{16\pi^2}\frac{1}{75600 m^4}(1217-4480 \delta_2)$ & $\frac{1}{16\pi^2}\frac{1}{75600 m^4}(1297-11760 \delta_2)$ \\ \hline
        $2$ & $\frac{1}{16\pi^2}\frac{1}{504 m^2}(-1\textcolor{Green}{+420 \delta_2})$ & $\frac{1}{16\pi^2}\frac{1}{7560 m^4}(2018\textcolor{Green}{+9051 \delta_2})$ & $\frac{1}{16\pi^2}\frac{1}{7560 m^4}(2008\textcolor{Green}{+10591 \delta_2})$ \\ \hline
    \end{tabular}
    \caption{
        \label{tab:GREFT_C34}
Wilson coefficients $C_3$, $C_{4,1}$, $C_{4,2}$ from matching the GREFT to UV completions with spinning particles with non-minimal gravitational interactions, up to spin $S=2$ and linear quadrupole terms. 
Quadrupole corrections not satisfying the calculability condition in~\cref{eq:EH_calculability} are highlighted in green to mark that other unknown corrections of the same order in the EFT expansion exist.} 
\end{table}
\renewcommand{\arraystretch}{1}

At this point we include the effects of the quadrupole coupling of matter to gravity.
According to \cref{eq:EH_calculability}, 
the linear quadrupole contributions to $C_{4,1}$ and $C_{4,2}$ ($n=4$) are calculable with our method for $S=1$ and $S=3/2$. 
On the other hand, such contributions to $C_3$ are never calculable beyond minimal coupling. 
The actual computation is not different in any way from the previous one, just slightly more involved.
In order to tackle this,  we implemented a program in Mathematica that contracts massive or massless spinors through xAct, with the package xSpinors \cite{martin-garcia_xact,Gomez-Lobo:2011kaw}.
Performing this way the phase space integration and $1/m$ expansion allows us to identify the contributions to Wilson coefficients summarized in \cref{tab:GREFT_C34}. 
There, for illustrative purpose, we also display the calculated  contributions that do not satisfy the condition in \cref{eq:EH_calculability}. 
 In these cases, highlighted in color, it should be understood that the results assume the absence of rational and tadpole terms contributions, and are expected to be of the same order as the contributions from unknown physics generating $\delta_2$. 
 Once again, Wilson coefficients of higher-dimensional operators can also be extracted from the amplitudes we constructed, 
 and selected results up to dimension 16 are displayed in \cref{tab:higher order coefficients GREFT} in \cref{app:higherorder}.  

\section{Positivity} 
\label{sec:positivity}

In the previous sections we collected new data relative to a class of UV completions for the theory of low energy photons or gravitons. 
Given such data, it is interesting to see whether these theories are in agreement with the existing positivity bounds. 
Our discussion follows closely the lines of  Ref.~\cite{Henriksson:2021ymi} for EHEFT and Ref.~\cite{Bern:2021ppb} for GREFT.

Positivity bounds select regions in the space of all Wilson coefficients of an EFT.
The positivity program consists in ruling out certain regions of this space using broad general assumptions of unitarity, crossing, and polynomial boundedness for the UV completion, while staying agnostic about all of its other features. 
In the literature it is customary to consider a space of ratios of couplings: there, the allowed points in theory space are convex regions known as convex hulls.

In the following, we will define higher-dimensional Wilson coefficients, for which the the structure of the positivity bounds is richer than for the ones we computed above. 
We will show how ratios of such coefficients move within the allowed regions as a function of the multipole coefficients. 
We will see how multipole effects move non-minimal (positive) theories within the space, with a general trend of moving towards "more positive" regions. 
In the case of GREFT, it was shown in Ref.~\cite{Bern:2021ppb} how known UV completions generate Wilson coefficients confined to islands that are much smaller than the region allowed by the existing positivity constraints. 
We will then see how non-minimal deformations of the theories are still strongly confined within such islands. 

\subsection{Constraints on EHEFT}

Let's first define the higher-order Wilson coefficients.
In order to do so, we could add a set of independent operators at each wanted order in the Lagrangian in~\cref{eq:INT_Leheft}. 
It is however more practical and convenient to define these directly at the amplitude level.
To this end, we add in \cref{eq:4-photons -- EH} and \cref{eq:4-photons -+ EH} terms with higher powers of the Mandelstram invariants that are invariant under the exchange of same-helicity bosons. Following Ref.~\cite{Henriksson:2021ymi} we define
\begin{align}
    &\cM_-= \cM\left(1_\gamma^-,2_\gamma^-,3_\gamma^-,4_\gamma^-\right)=
    \frac{\la12\ra^2\la34\ra^2}{s^2}\left[
    f_2\left(s^2+t^2+u^2\right)
    +f_3 stu 
    + f_4\left(s^2+t^2+u^2\right)^2
    +\ldots\right]
    ,\notag \\
&\cM_+=
\cM\left(1_\gamma^-,2_\gamma^-,3_\gamma^+,4_\gamma^+\right)=
\la12\ra^2[34]^2\left[
g_2 
+g_3 s
+ g_{4,1} s^2 
+ g_{4,2} (s^2+t^2+u^2)+\ldots\right],
    \label{eq:higher order Wilson EH}
\end{align}
where the subscript $k$ in each coefficient refers to a corresponding $2k+4$-dimensional operator with $2k$ derivatives in the Lagrangian, and $g_2=C_+$,  $f_2=C_-$.
The values of the Wilson coefficients defined above are collected in \cref{tab:higher order coefficients EH}, and they agree in the limit of minimal coupling with the values reported in Ref.~\cite{Henriksson:2021ymi}.

\begin{figure}[htb]
     \begin{subfigure}[]{0.49\textwidth}
         \centering
         \includegraphics[width=\textwidth]{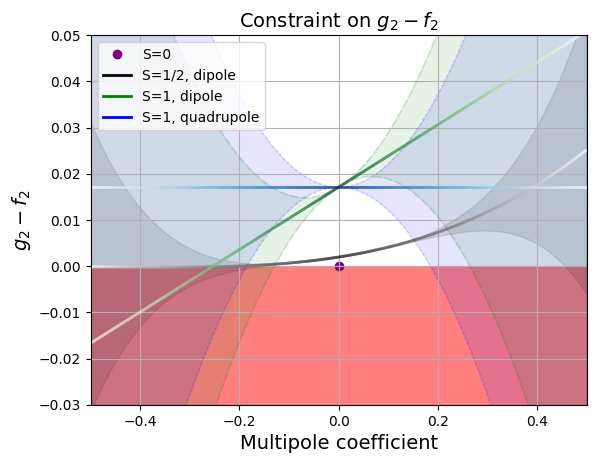}
     \end{subfigure}
     \hfill
     \begin{subfigure}[]{0.49\textwidth}
         \includegraphics[width=\textwidth]{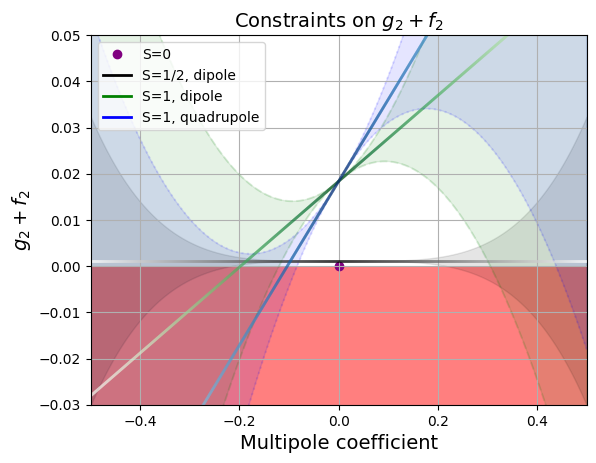}
     \end{subfigure}
        \caption{%
Positivity constraints on $g_2-f_2 = C_1$ (left) and $g_2 + f_2 = C_2$ (right) in EHEFT, and points in theory space matching to UV completions including massive spinning particles interacting non-minimally,  displayed in units of $1/m^4$. 
Positivity is violated, in the red region.
The envelope around the curves corresponds to an uncertainty due to contributions from degrees of freedom above the cutoff of the non-minimally coupled theory. }
        \label{fig:positivity of dim4 coefficients EH}
\end{figure}

Positivity constraints on dimension-8 coefficients read
\begin{equation}
    g_2 + f_2 > 0,\qquad \qquad 
    g_2 - f_2 > 0,
\end{equation}
and for the theories at hand they are shown, as a function of the multipole coefficient, in \cref{fig:positivity of dim4 coefficients EH}.
We remark that the curves in \cref{fig:positivity of dim4 coefficients EH} include terms up to cubic order in $\delta_1$ for $S=1/2$ and up to linear order in $\delta_1$  and $\delta_2$ for $S=1$, because higher order terms are not calculable. 
The plots are qualitatively showing that, for small enough values of the multipole coefficients, where it is safe to neglect orders beyond linear, positivity bounds remain satisfied. 
For greater values the positivity bounds may seemingly be violated. 
But in this regime both higher-order effects in $\delta_i$ and the contributions from the unknown UV completion above $\Lambda$ may easily push the curve back into the allowed region.  
This is in fact evident for $S=1/2$, where the (calculable) $\cO(\delta_i^2)$ make the curve asymptote toward the boundary $g_2 - f_2 =0$ of the allowed region. 
In~\cref{fig:positivity of dim4 coefficients EH} we illustrated the uncertainty due to unknown higher-order corrections in the EFT expansion by adding to the curves the error envolopes scaling as $1/\Lambda^4$ with $\Lambda \sim m/\delta^{1 \over 2S}$.

\begin{figure}
     \centering
     \begin{subfigure}[]{0.8\textwidth}
         \centering
         \includegraphics[width=\textwidth]{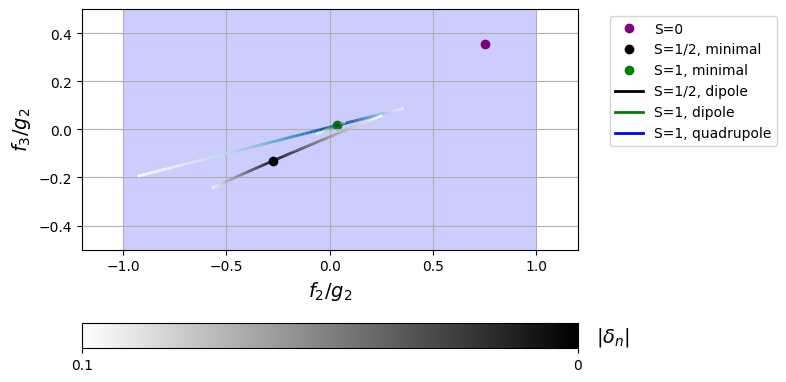}
         \caption{Constraints in  $\left[\frac{f_2}{g_2}, \frac{f_3}{g_2}\right]$ space.}
     \end{subfigure}
     \hfill
     \begin{subfigure}[]{0.8\textwidth}
         \centering
         \includegraphics[width=\textwidth]{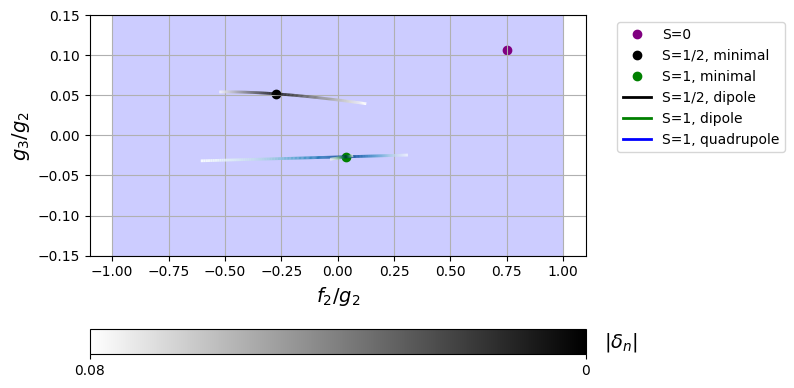}
         \caption{
         Constraints in  $\left[\frac{f_2}{g_2}, \frac{g_3}{g_2}\right]$ space.
         }
     \end{subfigure}
     \hfill
        \begin{subfigure}[]{0.8\textwidth}
         \centering
         \includegraphics[width=\textwidth]{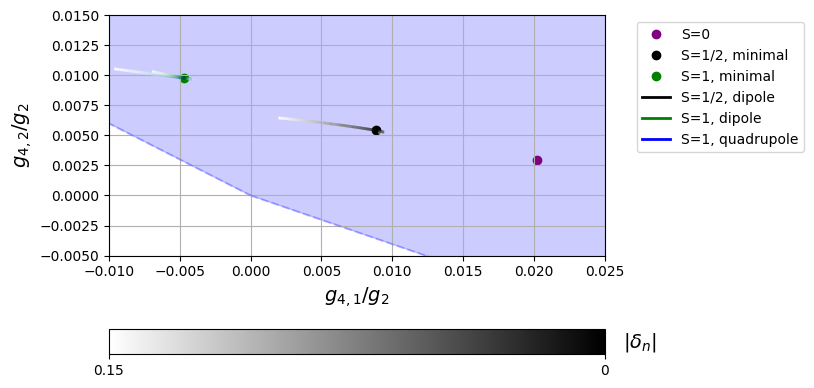}
         \caption{Constraints in  $\left[\frac{g_{4,1}}{g_2}, \frac{g_{4,2}}{g_2}\right]$ space.}
         \label{fig:positivity of dim8 coefficients EH}
     \end{subfigure}
    \caption{Positivity constraints on dimension-10 and -12 Wilson coefficients in EHEFT, and points in the theory space matching to UV completions including massive spinning particles interacting non-minimally. 
    The blue hulls corresponds to allowed theories. 
    Darker lines corresponds to smaller multipole coefficients, meaning that our analysis including up to linear corrections is more reliable there.}
        \label{fig:positivity of dim6 dim8 coefficients EH}
\end{figure}

Next, we study positivity constraints on dimension-10 and -12 Wilson coefficients, found numerically in Ref.~\cite{Henriksson:2021ymi} for minimal coupling, and shown in \cref{fig:positivity of dim6 dim8 coefficients EH}.
In the case of dimension-10 coefficients, we see how the bound on $f_2$ tends to get saturated (analogously to what is seen in \cref{fig:positivity of dim4 coefficients EH}), while the ones on $g_3$ and $f_3$ are weakly affected by multipole interactions.
Dimension-12 Wilson coefficients also mostly move along lines which are along the direction  the border of the hull.

\subsection{Constraints on GREFT}

In Ref.~\cite{Bern:2021ppb} it was shown that the Wilson coefficients arising from several UV completions of the GREFT lie on islands which are much smaller than the ones suggested by the sole constraints of unitarity, crossing, and Regge boundedness. 
The authors have explained the existence of these islands with a new constraint, Low Spin Dominance (LSD). 
While we refer to that reference for a detailed explanation, we here show how deformations of the considered theories via non-minimal quadrupole interactions remain confined to the same islands.

\begin{figure}[htb]
     \begin{subfigure}[]{0.49\textwidth}
         \centering
         \includegraphics[width=\textwidth]{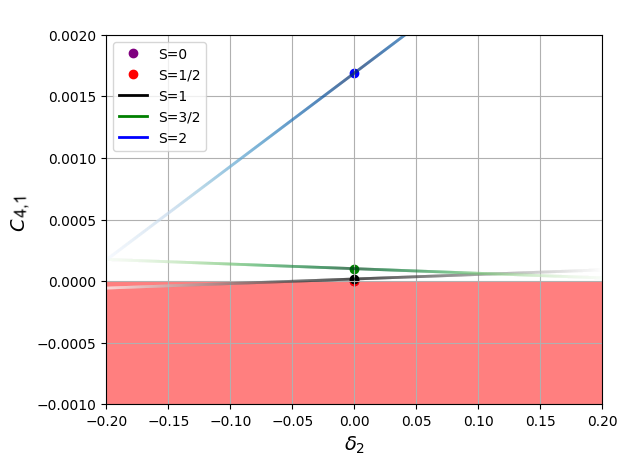}
     \end{subfigure}
     \hfill
     \begin{subfigure}[]{0.49\textwidth}
         \includegraphics[width=\textwidth]{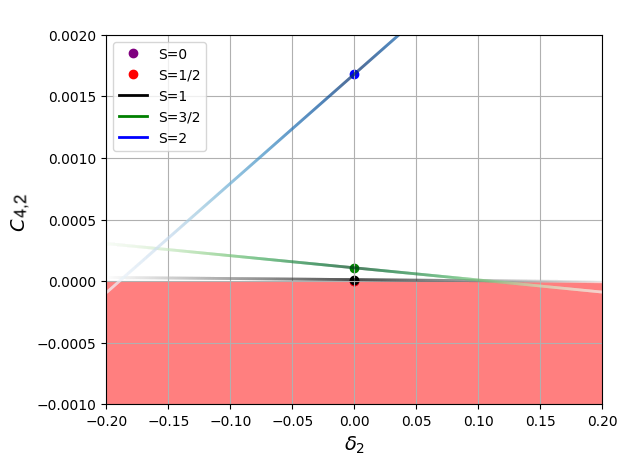}
     \end{subfigure}
        \caption{Positivity constraints on the dimension-8 Wilson coefficients $C_{4,1}$ and $C_{4,2}$ in GREFT, and points in theory space matching to UV completions including massive spinning particles  characterized by anomalous contributions to the quadrupole gravi-magnetic moment  proportional to $\delta_2$. 
        Positivity is violated in the red region.
   Darker lines corresponds to smaller quadrupole coefficients, meaning that our analysis including up to linear corrections is more reliable there. }
        \label{fig:positivity of dim4 coefficients GREFT}
\end{figure}

First, we show in \cref{fig:positivity of dim4 coefficients GREFT} how the dimension-8 Wilson coefficients $C_{4,1}$ and $C_{4,2}$ change as a function of the quadrupole.
Much as in EHEFT, these coefficients turn negative for large values of $|\delta_2|$ where the uncalcualable contributions become significant.

\begin{figure}[htb]
    \centering
    \includegraphics[width=0.8\linewidth]{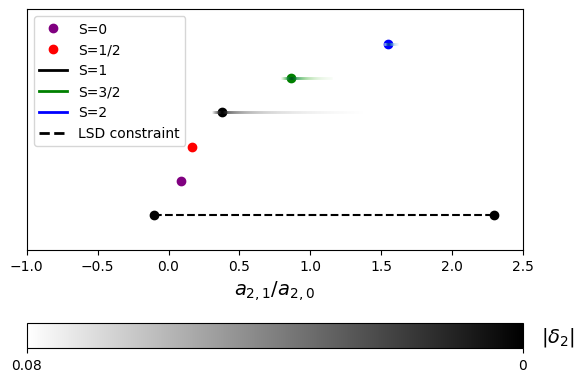}
    \caption{
    Points in the space of dimension-12 GREFT Wilson coefficients matching to UV completions with massive spinning particles characterized by anomalous contributions to the quadrupole gravi-magnetic moment proportional to $\delta_2$. 
Darker lines corresponds to smaller quadrupole coefficients, meaning that our analysis including up to linear corrections is more reliable there. 
    The dashed line represents the LSD constraint derived in Ref.~\cite{Bern:2021ppb}, while usual positivity bounds constrain the coefficients to a region roughly five times larger. }
    \label{fig:positivity of dim8 coefficients GREFT}
\end{figure}

Following Ref.~\cite{Bern:2021ppb}, we now define higher order Wilson coefficients, by parametrizing the contact part of the MHV four-graviton amplitude as
\begin{align}
\cM[1_h^-,2_h^-,3_h^+,4_h^+]=\frac{\la12\ra^4[34]^4}{\mpl^4}\sum_{k\geq j\geq0} a_{k,j} s^{k-j} t^j,
    \label{eq:higher order Wilson GREFT}
\end{align}
where crossing gives relations among several $a_{k,j}$ parameters and, in particular, $a_{0,0}=C_+$.
Positivity bounds constrain the dimension-12 Wilson coefficients $a_{2,k}$ as\footnote{Notice that crossing relates $a_{2,1}=a_{2,2}$.}
\begin{equation}
    -\frac{90}{11}\leq\frac{a_{2,1}}{a_{2,0}}\leq 6,
\end{equation}
but  all the minimal coupling data from Ref.~\cite{Bern:2021ppb} and the deformations we provided lie in the much smaller region shown in \cref{fig:positivity of dim8 coefficients GREFT}.
All points lie within the LSD bounds, and quadrupole corrections shift them only slightly.
The largest shifts  happens to be in a direction where the bound is more comfortably satisfied.

\begin{figure}[htb]
    \centering
    \includegraphics[width=0.8\linewidth]{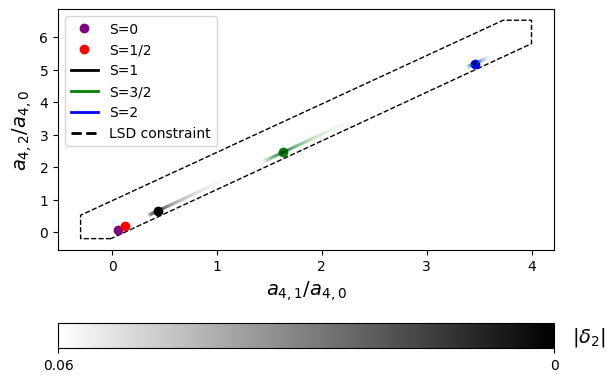}
    \caption{%
    Points in the space of dimension-16 GREFT Wilson coefficients matching to UV completions with massive spinning particles characterized by anomalous contributions to the quadrupole gravi-magnetic moment proportional to $\delta_2$.
Darker lines corresponds to smaller quadrupole coefficients, meaning that our analysis including up to linear corrections is more reliable there. 
The dashed line represents the LSD constraint~\cite{Bern:2021ppb}, while the usual positivity bounds constrain the coefficients to a much larger region that would cover all the space shown.
The slope of the fading lines is $\mathcal{O}\left(10^{-4}\right)$ away from the most stringent LSD constraint, which collapses on the line with slope $3/2$.}
    \label{fig:positivity of dim12 coefficients GREFT}
\end{figure}
The confinement to small islands is more spectacular for dimension-16 Wilson coefficients. 
This is illustrated  in \cref{fig:positivity of dim12 coefficients GREFT} in the plane $\left[\frac{a_{4,1}}{a_{4,0}};\frac{a_{4,2}}{a_{4,0}}\right]$.
The dashed region again refers to the LSD constraint, and is much smaller than the one derived using positivity constraints. All the  data lies within the region. 
Interestingly, the slope of the deformation approaches the slope of the LSD  region, $\frac{3}{2}$, for all spins.

\begin{figure}[htb]
    \centering
    \includegraphics[width=0.8\linewidth]{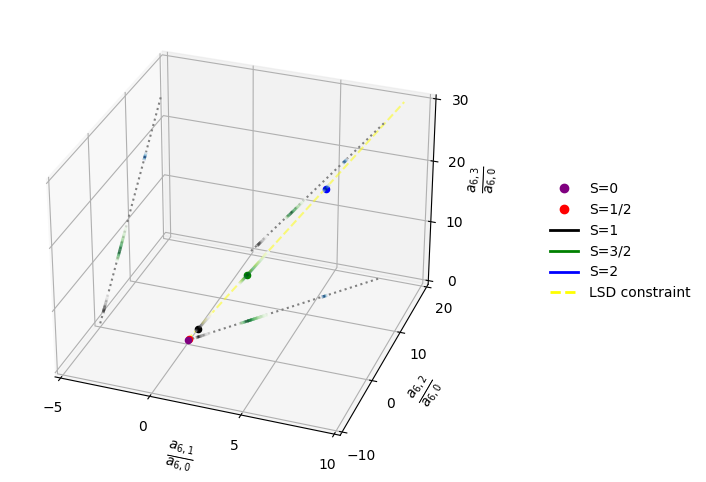}
    \caption{%
 Points in the space of dimension-20 GREFT Wilson coefficients matching to UV completions with massive spinning particles characterized by anomalous contributions to the quadrupole gravi-magnetic moment proportional to $\delta_2$.
 Darker lines corresponds to smaller quadrupole coefficients, meaning that our analysis including up to linear corrections is more reliable there.
The yellow line represents the LSD constraint~\cite{Bern:2021ppb}, while usual positivity bounds constrain the coefficients to a much larger region. 
Dashed gray lines and colored fading lines on top of them are projection of the 3D curve on each of the displayed planes.}
    \label{fig:positivity of dim16 coefficients GREFT}
\end{figure}

Finally we show, in figure \cref{fig:positivity of dim16 coefficients GREFT}, how quadrupole corrections move dimension-20 Wilson coefficients in the theory space. 
Again, the coefficients are allowed to be deformed only along the region resulting from applying the LSD constraint. 
The allowed theory space shrinks along a line in this higher-dimensional case as well. 
Once again, the quadrupole corrections align perfectly within the narrow  passageway defined by the theory space in the limit of minimal coupling.

\section{Conclusions} 
\label{sec:conclusions}

In this paper we have calculated the Wilson coefficients in the low-energy theory of photons and gravitons, matching it at one loop to another EFT with a spinning matter particle. 
The novelty is that matter is allowed to have general electromagnetic and gravitational interactions, including non-minimal multipoles.  
The complexities of such a calculation can be eased with a help of the on-shell amplitude formalism. 
Relying on the principles of unitarity and locality, we constructed the one-loop 4-photon helicity amplitudes with up to spin-1 matter in the loop having an anomalous magnetic dipole and quadrupole moment. 
In our approach, only two-particle cuts in four dimensions are used to glue two Compton amplitudes and reconstruct the box, triangle, and bubble coefficients of the loop amplitude. 
Rational terms can then be uniquely fixed in a number of cases by requiring absence of non-local poles. 
The same approach was applied to the 4-graviton amplitudes with up to spin-2 matter with the anomalous gravitational quadrupole moment.
The only technical difference is that in this case tadpoles are also needed to cancel non-local divergences, and that a divergent contribution corresponding to renormalization of the Planck mass cannot be determined by our method. 

Expanding the analytic expressions for the amplitudes at large values of the matter particle mass, one can efficiently extract the EHEFT and GREFT Wilson coefficients at arbitrary dimensions. 
The results are summarized in \cref{eq:EH_C12-fermion,eq:EH_C12-vector,tab:higher order coefficients EH,tab:GREFT_C34,tab:higher order coefficients GREFT}. 
We also give a compact criterion in \cref{eq:EH_calculability} as to whether the multipole corrections to the Wilson coefficients are calculable, that is whether they are parametrically larger than generic effects induced by heavy physics above the cutoff of the matter EFT.   

One goal of this exercise was to explore the theory space of EHEFT and GREFT. 
In both cases, only the leading terms in the Lagrangian ($F_{\mu\nu}^2$ and $\mpl^2 R$, respectively) have been experimentally observed.  
There is an ongoing program to pinpoint or constrain the subleading interaction terms in these effective theories. 
Our results provide a more general benchmark to interpret the future findings. 
Another goal was purely theoretical. 
In the recent years the program of inspecting the parameter space of EFTs using the so-called positivity constraints has received a lot of attention. 
For EHEFT and especially for GREFT, it was found that the theory space allowed by positivity is much larger than the regions accessible by known UV completions~\cite{Henriksson:2021ymi,Bern:2021ppb}. 
Our results largely confirm these conclusions, showing that multipole corrections remain confined to small islands in the theory space. 
This is perhaps most spectacularly seen in \cref{fig:positivity of dim12 coefficients GREFT,fig:positivity of dim16 coefficients GREFT}, where gravitational quadrupole corrections move the Wilson coefficients along narrow passageways determined by the previous minimal coupling results.

\section*{Acknowledgements}
AF has received funding from the Agence Nationale de la Recherche (ANR) under grant ANR-19-CE31-0012 (project MORA) and from the European Union’s Horizon 2020 research and innovation programme under the Marie Sklodowska-Curie grant agreement No 860881-HIDDeN. 

\appendix

\newpage
\section{Integrals} 
\label{app:INT}

In this appendix we discuss the dictionary between two-particle cut discontinuities of the one-loop amplitude and the discontinuities of the standard basis of scalar integrals.

\subsection{Parametrization}
\label{App: parametrization}

First of all, we define the variables used in \cref{sec:EH,sec:GREFT} to parametrize the on-shell phase space of two-particle intermediate states in the cut 4-photon and 4-graviton amplitudes.
In the following, this parametrization is defined for $s$-channel discontinuities; the one for the $t$-channel is straightforwardly found by crossing $(1\leftrightarrow3)$.

For $s$-channel cuts, the massive internal momenta $p_X$, $p_Y$, 
$p_X^2 = p_Y^2 = m^2$ 
are decomposed as
\begin{align}
    &p_X=\alpha p_1+(1-\alpha) p_2-\sqrt{\alpha(1-\alpha)-\frac{m^2}{s}}[zq+z^{-1} \bar q],\notag\\
    &p_Y=(1-\alpha) p_1+\alpha p_2+\sqrt{\alpha(1-\alpha)-\frac{m^2}{s}}[zq+z^{-1} \bar q]
    \label{eq: parametrization s channel}
, \end{align}
where $p_1$, $p_2$ are the incoming massless momenta, $p_1^2=p_2^2=0$.
The other two basis momenta are defined as 
$q \sigma = |2\rangle [1|$, 
$\bar q \sigma = |1\rangle [2|$, 
such that $q^2 = \bar q^2 = 0$ and 
$2 q \bar q = - s$. 
The parameter $\alpha$ is in the range 
\begin{equation}
\alpha\in[\alpha_-,\alpha_+],\qquad\alpha_\pm=\frac{1\pm\sqrt{1-\frac{4m^2}{s}}}{2},
\end{equation}
and the other parameter is constrained to the unit circle, $|z|=1$
In the limit $m \to 0$ this reduces to the parametrization introduced in Refs.~\cite{Zwiebel:2011bx,Caron-Huot:2016cwu} with $\alpha =\cos^2 \theta$, $z=e^{i\phi}$.  
In these variables, 
the  phase space element of the two massive particle becomes 
\begin{align}
\label{eq:INT_phasespace}
\text{d}\Pi_{XY}=
\frac{\text{d}\alpha}{8\pi}
\frac{\text{d}z}{2\pi i z}
.\end{align}

\subsection{Scalar integrals}
On general grounds,
a one-loop amplitude can be decomposed as  
\begin{align}
    \cM=\sum_i c_\circ^i I_\circ^i+\sum_i c_\triangleright^i I_\triangleright^i+\sum_i c_\Box^i I_\Box^i+c_1 \Delta I +R 
\end{align}
where $I_n^i$ are the tadpole, bubble, triangle, and box scalar integrals, 
$\Delta I$ marks contributions of tadpoles $I_1$ and massless bubbles, and $R$ stands for rational terms.

For our purposes we only need to consider scalar integrals with propagators carrying the same mass $m$. Thus, the basis of scalar integrals we pick comprises the following ten integrals:
\begin{align}
\label{APP:scalar integrals}
    &I_1=\int\frac{\text{d}^4 k}{i(2\pi)^4}\frac{1}{[k^2-m^2]},\notag\\
    &I_\circ^s=\int\frac{\text{d}^4 k}{i(2\pi)^4}\frac{1}{[k^2-m^2][(k+p_1+p_2)^2-m^2]},\notag\\
    &I_\circ^t=\int\frac{\text{d}^4 k}{i(2\pi)^4}\frac{1}{[k^2-m^2][(k+p_1+p_3)^2-m^2]},\notag\\
    &I_\circ^u=\int\frac{\text{d}^4 k}{i(2\pi)^4}\frac{1}{[k^2-m^2][(k+p_1+p_4)^2-m^2]},\notag\\
    &I_\triangleright^s=\int\frac{\text{d}^4 k}{i(2\pi)^4}\frac{1}{[k^2-m^2][(k+p_1+p_2)^2-m^2][(k+p_1)^2-m^2]},\notag\\
    &I_\triangleright^t=\int\frac{\text{d}^4 k}{i(2\pi)^4}\frac{1}{[k^2-m^2][(k+p_1+p_3)^2-m^2][(k+p_1)^2-m^2]},\notag\\
    &I_\triangleright^u=\int\frac{\text{d}^4 k}{i(2\pi)^4}\frac{1}{[k^2-m^2][(k+p_1+p_4)^2-m^2][(k+p_1)^2-m^2]},\notag\\
    &I_\Box^{st}=\int\frac{\text{d}^4 k}{i(2\pi)^4}\frac{1}{[k^2-m^2][(k+p_1)^2-m^2][(k+p_1+p_2)^2-m^2][(k-p_3)^2-m^2]},\notag\\
    &I_\Box^{su}=\int\frac{\text{d}^4 k}{i(2\pi)^4}\frac{1}{[k^2-m^2][(k+p_1)^2-m^2][(k+p_1+p_2)^2-m^2][(k-p_4)^2-m^2]},\notag\\
    &I_\Box^{tu}=\int\frac{\text{d}^4 k}{i(2\pi)^4}\frac{1}{[k^2-m^2][(k+p_1)^2-m^2][(k+p_1+p_3)^2-m^2][(k-p_4)^2-m^2]},
\end{align}
where the superscripts denote in which kinematic channels each integral has a discontinuity. 
The tadpoles and bubbles are implicitly dimensionally regularized with $d=4-2\epsilon$.
The discontinuities of such integrals are to be matched, at the integrand level, with the ones of the general four-bosons loop amplitude. In order to do so, we show how such discontinuities are computed with an example.

The box integrals are the most involved, so we'll present the case of $I_\Box^{su}$. Computing the discontinuity over the s-channel cut is equivalent to putting the propagators that mediate the s-channel exchange on the mass-shell:
\begin{align}
    \text{Disc}^s I_\Box^{su}=\int\frac{\text{d}^4 k}{i(2\pi)^2}\frac{\delta[k^2-m^2]\delta[(k+p_1+p_2)^2-m^2]}{(2k p_1) (2kp_4)}.
\end{align}
The numerator is proportional to the two-massive-particle phase space element.
Then, using \cref{eq: parametrization s channel,eq:INT_phasespace} we get
\begin{align}
    \text{Disc}^s I_\Box^{su}=&
    \frac{i}{8\pi s \la23\ra [13]}
\int_{\alpha_-}^{\alpha_+}\frac{\text{d}\alpha}{(1-\alpha)\sqrt{\alpha(1-\alpha)-\frac{m^2}{s}}}\int\frac{\text{d}z}{2\pi i}\frac{1}{[z-z_{s+}][z-z_{s-}]}
    \notag\\
    &=
    -\frac{i}{8\pi s u} 
\int_{\alpha_-}^{\alpha_+}\frac{\text{d}\alpha}{(1-\alpha)\sqrt{[\alpha-x_s(1-\alpha)]^2+\frac{x_s}{y_s}}},
\label{eq: discontinuity of su box example}
\end{align}
where
$x_s \equiv t/u$,
$y_s \equiv s/4m^2$, and
$z_{s_\pm}$ 
are defined in \cref{eq: zplusminus s channel}, and in the last passage we computed the $z$-integral using the residue theorem, noticing that that only $z_{s_-}$ is inside the unit circle.
The integral over $\alpha$ can also be performed but the result is not particularly revealing, and in fact it is much easier to match discontinuities at the level of $\alpha$ integrands.

\subsection{Dictionary}
\label{APP: dictionary}

Using the methods described in \cref{sec:EH}, we are able to reduce the $s$-channel discontinuities of one-loop amplitudes of interest to a form 
\begin{align}
\text{Disc}^s {\cal M} =   
i\int\frac{\text{d}\alpha}{\pi}
\bigg \{ &
 \frac{A_1}
 {\alpha\sqrt{[\alpha-x_s(1-\alpha)]^2+\frac{x_s}{y_s}}}
 + 
  \frac{A_2}
 {(1-\alpha)\sqrt{[\alpha-x_s(1-\alpha)]^2+\frac{x_s}{y_s}}}
\nnl  
+ & \sum_{n \geq 0} 
  \frac{B_n \alpha^n}
 {\sqrt{[\alpha-x_s(1-\alpha)]^2+\frac{x_s}{y_s}}}
 + \frac{C_1}{\alpha} + \frac{C_2}{1-\alpha}
+ \sum_{n \geq 0}  D_n \alpha^n
\bigg \} 
,\end{align}
where all integrals in this subsection are evaluated from 
$\alpha_-$ to $\alpha_+$. 
In order to translate that into a sum of discontinuities of the scalar integrals in \cref{APP:scalar integrals}, we use the following dictionary:
\begin{align}
 &i\int\frac{\text{d}\alpha}{\pi}
 \frac{1}
 {\alpha\sqrt{[\alpha-x_s(1-\alpha)]^2+\frac{x_s}{y_s}}}
 =   -8su\text{Disc}^s I_\Box^{st}
,    \notag\\
    &i\int\frac{\text{d}\alpha}{\pi}
    \frac{1}
    {(1-\alpha)\sqrt{[\alpha-x_s(1-\alpha)]^2+\frac{x_s}{y_s}}}
=-8s u\text{Disc}^s I_\Box^{su}
, \end{align}
\begin{align}
    &i\int\frac{\text{d}\alpha}{\pi}\frac{1}{\sqrt{[\alpha-x_s(1-\alpha)]^2+\frac{x_s}{y_s}}}
    =8u\text{Disc}^s I_\triangleright^{s}
  ,  \notag\\
    &i\int\frac{\text{d}\alpha}{\pi}\frac{\alpha}{\sqrt{[\alpha-x_s(1-\alpha)]^2+\frac{x_s}{y_s}}}=
    -\frac{8 t u}{s}
    \text{Disc}^s I_\triangleright^s
    -\frac{8u(t-u)}{s^2}
    \text{Disc}^s I_\circ^s
 ,    \notag\\
    &i\int\frac{\text{d}\alpha}{\pi}\frac{\alpha^2}
    {\sqrt{[\alpha-x_s(1-\alpha)]^2+\frac{x_s}{y_s}}}
    =\frac{8tu(st-2m^2 u)}{s^3}
    \text{Disc}^s I_\triangleright^s
    +\frac{4u(6t^2+2st-s^2)}{s^3}
    \text{Disc}^s I_\circ^s
, \end{align}
\begin{align}
    &i\int\frac{\text{d}\alpha}{\pi}\frac{1}{\alpha}
    =i\int\frac{\text{d}\alpha}{\pi}\frac{1}{1-\alpha}
    =-8 s \text{Disc}^s I_\triangleright^s
, \end{align}
\begin{align}
    &i\int\frac{\text{d}\alpha}{\pi}
    =8\text{Disc}^s I_\circ^s
  ,  \notag\\
    &i\int\frac{\text{d}\alpha}{\pi}\alpha
    =4\text{Disc}^s I_\circ^s
  ,  \notag\\
    &i\int\frac{\text{d}\alpha}{\pi}\alpha^2
    =\frac{8y_s-2}{3y_s}\text{Disc}^s I_\circ^s
  ,  \notag\\
    &i\int\frac{\text{d}\alpha}{\pi}\alpha^3
    =\frac{2y_s-1}{y_s}\text{Disc}^s I_\circ^s
  ,  \notag\\
    &i\int\frac{\text{d}\alpha}{\pi}\alpha^4
    =\frac{1-12y_s+16y_s^2}{10y_s^2}\text{Disc}^s I_\circ^s
   ,  \notag\\
    &i\int\frac{\text{d}\alpha}{\pi}\alpha^5
    =\frac{3-16y_s+16y_s^2}{12y_s^2}\text{Disc}^s I_\circ^s
. \end{align}
The $t$-channel dictionary can be obtained by crossing the $s$-channel one.

\section{Supplementary results} 
\label{app:RESULT}

\subsection{Compton amplitudes}

We present here more general results for the same-sign gravi-Compton amplitudes involving a (complex) massive particle $X$ interacting non-minimally with the graviton. 
Up to linear order in multipole interactions we can split the amplitude as 
\begin{align}
\cM\left(1_h^- 2_h^- \bold 3_{X} \bold 4_{\bar\phi}\right) = 
\cM_0  + \delta_2 \cM_2 + \delta_3 \cM_3 + \delta_4 \cM_4 + \cO(\delta_n^2)  
. \end{align}
We give the amplitude up to contact terms.
In the limit of minimal coupling  
\begin{align}
\cM_0 = 
 \frac{ \la 12\ra^4 
 \mls{3}\mrs{4}^{2S}  }
 {\mpl^2 m^{2S-4} s (t-m^2) (u-m^2)} 
, \end{align}
which is valid for any $S$. 
The linear quadrupole corrections for 
$S \geq 1$ read 
\begin{align}
\cM_2 = &
 \frac{ \la 12\ra^4 
  \mls{4}\mrs{3}^{2S-2} 
  \big (\mls{4}\mrs{3} -\mla{4}\mra{3} \big )^2}
 {\mpl^2 m^{2S-4} s (t-m^2) (u-m^2)} 
\nnl - & 
 \frac{\la 12\ra^2  \mls{4}\mrs{3}^{2S-2} }  {\mpl^2 m^{2S-2}} 
\bigg \{ 
2 { \la1\mra{3} \la2\mra{4} \la1\mra{4} \la2\mra{3}
\over (t-m^2) (u-m^2)}
+  { \la1\mra{3}^2 \la2\mra{4}^2 
\over m^2 (t-m^2)}
+  { \la1\mra{4}^2  \la2\mra{3}^2 
\over m^2 (u-m^2)}
\bigg \} 
.\end{align}
The linear octupole corrections for 
$S \geq 3/2$ read 
\begin{align}
\cM_3 = &
 \frac{ \la 12\ra^4 
  \mls{4}\mrs{3}^{2S-3} 
  \big (\mls{4}\mrs{3} -\mla{4}\mra{3} \big )^3}
 {\mpl^2 m^{2S-4} s (t-m^2) (u-m^2)} 
 - 
  \frac{\mls{4}\mrs{3}^{2S-3} } 
  {\mpl^2 m^{2S-2}}
  \bigg \{ 
{ 3 \la 12\ra^2 \la1\mra{3} \la2\mra{4} \la1\mra{4} \la2\mra{3} 
 \big (\mls{4}\mrs{3} -\mla{4}\mra{3} \big )
\over (t-m^2) (u-m^2)}
 \nnl  &   
+  { 3 \la 12\ra \la1\mra{3}^2 \la2\mra{4}^2 
\la1\mra{4} \la2\mra{3} 
\over m^2 (t-m^2)}
- { \la 12\ra^2 \la1\mra{3}^2 \la2\mra{4}^2 
\big ( \mla{4}| p_1 |\mrs{3} + \mla{3} |p_1 |\mrs{4} \big )  
\over m^3 (t-m^2)}
\nnl & 
- { 3 \la 12\ra \la1\mra{4}^2 \la2\mra{3}^2 
\la1\mra{3} \la2\mra{4} 
\over m^2 (u-m^2)}
- { \la 12\ra^2 \la1\mra{4}^2 \la2\mra{3}^2 
\big ( \mla{4}| p_2 |\mrs{3} + \mla{3} |p_2 |\mrs{4} \big )  
\over m^3 (u-m^2)}
  \bigg \} 
 .\end{align}
 The linear hexadecapole corrections for 
$S \geq 2$ read 
\begin{align}
\cM_4 = &
   {    \langle 1 2 \rangle^4  \mls{4}\mrs{3}^{2S-4}   \big (\mls{4}\mrs{3} -\mla{4}\mra{3} \big )^4
    \over m^{2S-4 } \mpl^2  s (t-m^2) (u-m^2)}    
 - { 4 \langle 1 2 \rangle^2  
\mls{4}\mrs{3}^{2S-4}   
\la1\mra{3} \la2\mra{4} \la1\mra{4}  \la2\mra{3} 
 \big ( \mla{4}| p_1 |\mrs{3}^2  
 +  \mla{4}| p_2 |\mrs{3}^2  \big )
 \over 
 m^{2S } \mpl^2 (t-m^2) (u-m^2)}   
   \nnl  \hspace{-2.5cm} +  & 
  \bigg ( 
  {  \mls{4}\mrs{3}^{2S-4}   
  \la1\mra{3}^2  \la2\mra{4}^2 
    \over m^{2S } \mpl^2 (t-m^2)  }    \bigg \{
6  \la1\mra{4}^2 \la2\mra{3}^2 
+ { 4 \langle 1 2 \rangle
\la1\mra{4} \la2\mra{3}   
\over m} 
 \bigg [ 
 2 \mla{4}| p_2 |\mrs{3}
 + 3 \mla{4}| p_1 |\mrs{3} 
-3 \mla{3}| p_2 |\mrs{4}   \bigg ] 
         \nnl  \hspace{-2.5cm} &   
- {  \langle 1 2 \rangle^2    \over m^2} 
       \bigg [  
\big ( \mla{4}| p_1 |\mrs{3} 
+\mla{3}| p_1 |\mrs{4}  \big )^2  
+16 \mla{4}| p_1 |\mrs{3} \mla{3}| p_2 |\mrs{4}  \bigg ]  
 \bigg \}   
 + (1 \leftrightarrow 2) \bigg ) 
  .\end{align}
 Note that, in the presence of multipole corrections, the same-sign amplitude behaves as 
 $\cM \sim \delta_n E^{2S+n}/(m^{2S+n-2}\mpl^2)$, 
 indicating the power counting 
 $\delta_n \sim (m/\mpl)^{2S+n-2}$.

\subsection{X-functions}
\label{APP: Result X functions}

Here we give intermediate results for the calculation of $t$-channel discontinuities of the MHV 4-photon and 4-graviton one-loop amplitudes   with minimally-coupled matter of spin $S$ in the loop.
After some manipulations, they can be brought to the following form 
\begin{equation}
 \text{Disc}^t  \cM(1_x^- 2_x^- 3_x^+ 4_x^+) =
    i \la12\ra^{2|h_x|}[34]^{2|h_x|}   
    \int\text{d}\Pi_{XY}\frac{X^{(S)}(z,\alpha)}
    {(2p_1 p_X) ( 2p_2 p_X)( 2p_3 p_X) (2p_4 p_X)}
, \end{equation}
where $x =\gamma(h)$ and $|h_x|=1(2)$ for photons (gravitons). 
For photons we have 
 \begin{align}
 \label{eq:RESULTS_XfunctionsPhotons}
X^{(0)} =  &  
 4 q_X^4 (\lambda_1 p_X \sigma \tilde \lambda_3)^2 (\lambda_2 p_X \sigma \tilde \lambda_4)^2 
,\nnl 
X^{(\frac{1}{2})} =  &  
 4 q_X^4 (\lambda_1 p_X \sigma \tilde \lambda_3) (\lambda_2 p_X \sigma \tilde \lambda_4)
\bigg \{  t  \langle 12 \rangle [34]   
- 2  (\lambda_1 p_X \sigma \tilde \lambda_3) (\lambda_2 p_X \sigma \tilde \lambda_4) 
\bigg \} 
,\nnl 
X^{(1)} =  &  
4 q_X^4  \bigg \{  \big ( t  \langle 12 \rangle [34] \big)^2 
 -4  t  \langle 12 \rangle [34]   (\lambda_1 p_X \sigma \tilde \lambda_3) (\lambda_2 p_X \sigma \tilde \lambda_4) 
 +3   (\lambda_1 p_X \sigma \tilde \lambda_3)^2 (\lambda_2 p_X \sigma \tilde \lambda_4)^2    
 \bigg \} 
, \end{align}  
while for gravitons we have
\begin{align}
\label{eq:RESULTS_XfunctionsGravity}
 & X^{(0)}=
  { (\lambda_1 p_X \sigma \tilde \lambda_3)^4(\lambda_2 p_X \sigma \tilde \lambda_4)^4 
  \over
   \mpl^4  t^2  \langle 1 2 \rangle^4 [34]^4  }   
 , \notag\\
   & X^{(\frac{1}{2})}=
     { (\lambda_1 p_X \sigma \tilde \lambda_3)^3(\lambda_2 p_X \sigma \tilde \lambda_4)^3 
  \over
   \mpl^4  t^2  \langle 1 2 \rangle^4 [34]^4  }   
   \bigg \{      
    t \langle 12 \rangle [34] - 2    (\lambda_1 p_X \sigma \tilde \lambda_3) (\lambda_2 p_X \sigma \tilde \lambda_4) 
             \bigg \} 
  , \notag\\
   & X^{(1)}=
     { (\lambda_1 p_X \sigma \tilde \lambda_3)^2(\lambda_2 p_X \sigma \tilde \lambda_4)^2
  \over
   \mpl^4  t^2  \langle 1 2 \rangle^4 [34]^4  }   
   \bigg \{    
   t^2 \langle 12 \rangle^2 [34]^2  - 4   t   \langle 12 \rangle  [34]  (\lambda_1 p_X \sigma \tilde \lambda_3) (\lambda_2 p_X \sigma \tilde \lambda_4) 
   \nnl &  + 3  (\lambda_1 p_X \sigma \tilde \lambda_3)^2 (\lambda_2 p_X \sigma \tilde \lambda_4)^2   
            \bigg \} 
  , \notag\\
   & X^{(\frac{3}{2})}=
    { (\lambda_1 p_X \sigma \tilde \lambda_3) (\lambda_2 p_X \sigma \tilde \lambda_4)
  \over
   \mpl^4  t^2  \langle 1 2 \rangle^4 [34]^4  }   
   \bigg \{    
    t^3 \langle 12 \rangle^3 [34]^3  
  - 6    t^2   \langle 12 \rangle^2  [34]^2  (\lambda_1 p_X \sigma \tilde \lambda_3) (\lambda_2 p_X \sigma \tilde \lambda_4) 
   \nnl & 
     + 10 t  \langle 12 \rangle  [34]  (\lambda_1 p_X \sigma \tilde \lambda_3)^2 (\lambda_2 p_X \sigma \tilde \lambda_4)^2 
  -4(\lambda_1 p_X \sigma \tilde \lambda_3)^2 (\lambda_2 p_X \sigma \tilde \lambda_4)^2 
         \bigg \} 
 ,  \notag\\
   & X^{(2)} =
 { 1  \over
   \mpl^4  t^2  \langle 1 2 \rangle^4 [34]^4  }   
   \bigg \{   
     t^4 \langle 12 \rangle^4 [34]^4
- 8 t^3 \langle 12 \rangle^3 [34]^3  (\lambda_1 p_X \sigma \tilde \lambda_3) (\lambda_2 p_X \sigma \tilde \lambda_4) 
 \nnl  &
  +21   t^2   \langle 12 \rangle^2  [34]^2  (\lambda_1 p_X \sigma \tilde \lambda_3)^2 (\lambda_2 p_X \sigma \tilde \lambda_4)^2
     - 20 t  \langle 12 \rangle  [34]  (\lambda_1 p_X \sigma \tilde \lambda_3)^3 (\lambda_2 p_X \sigma \tilde \lambda_4)^3
      \nnl  &
  + 5 (\lambda_1 p_X \sigma \tilde \lambda_3)^4 (\lambda_2 p_X \sigma \tilde \lambda_4)^4
         \bigg \} 
.\end{align}
The numerical  coefficients in the curly brackets in \cref{eq:RESULTS_XfunctionsPhotons,eq:RESULTS_XfunctionsGravity} are in one-to-one correspondence with the supersymmetric decomposition of the amplitudes~\cite{Dunbar:1994bn,Bern:2021ppb}. 
Given this form, it is straightforward to apply the parametrization in \cref{App: parametrization} and evaluate the contour integral over $z$, as described in more detail in \cref{sec:EH_minimal}. 
Then,  using the dictionary in 
\cref{APP: dictionary}, one can express the $t$-channel discontinuity as a sum of discontinuities of the master integrals in \cref{APP:scalar integrals}. 
Together with the discontinuity in the $u$ channel trivially obtained by crossing, and that in the $s$ channel computed by similar means, this allows one to reconstruct the box, triangle, and bubble coefficients in the decomposition of the MHV amplitude.  

\subsection{Higher-order Wilson coefficients}
\label{app:higherorder}

We here report all the results for dimension-6 and  -8 Wilson coefficients in EHEFT, and dimension-12, -16, and -20 Wilson coefficients in GREFT.
The results for EHEFT are shown in \cref{tab:higher order coefficients EH}, while the ones for GREFT are in \cref{tab:higher order coefficients GREFT}

\begin{table}[ht]
\centering
\begin{tabular}{|c|c|c|c|}
\hline
       & S=0 & S=1/2 & S=1 \\
\hline
$f_2$     & $\frac{e^4}{480 \cdot \pi^2}$ & $-\frac{e}{240 \cdot \pi^2} \left(e^3 + 20e^2\delta_1 + 50e\delta_1^2 + 40\delta_1^3\right)$ & $\frac{e}{160 \cdot \pi^2} \left(e^3 + 20e^2\delta_1 + 140e^2\delta_2\right)$ \\
\hline
$g_2$     & $\frac{e^4}{360 \cdot \pi^2}$ & $\frac{e}{720 \cdot \pi^2} \left(11e^3 + 60e^2\delta_1 + 150e\delta_1^2 + 120\delta_1^3\right)$ & $\frac{e}{240 \cdot \pi^2} \left(42e^3 + 190e^2\delta_1 + 210e^2\delta_2\right)$ \\
\hline
$f_3$     & $\frac{e^4}{1008 \cdot \pi^2}$ & $-\frac{e}{2520 \cdot \pi^2} \left(5e^3 + 84e^2\delta_1 + 231e\delta_1^2 + 210\delta_1^3\right)$ & $\frac{e}{1680 \cdot \pi^2} \left(5e^3 + 84e^2\delta_1 + 336e^2\delta_2\right)$ \\
\hline
$g_3$     & $\frac{e^4}{3360 \cdot \pi^2}$ & $\frac{e}{1260 \cdot \pi^2} \left(e^3 + 7e^2\delta_1 + 7e\delta_1^2 - 14\delta_1^3\right)$ & $-\frac{e}{10080 \cdot \pi^2} \left(47e^3 + 168e^2\delta_1 + 168e^2\delta_2\right)$ \\
\hline
$f_4$     & $\frac{e^4}{30240 \cdot \pi^2}$ & $-\frac{e}{60480 \cdot \pi^2} \left(4e^3 + 54e^2\delta_1 + 135e\delta_1^2 + 108\delta_1^3\right)$ & $\frac{e}{20160 \cdot \pi^2} \left(2e^3 + 27e^2\delta_1 + 85e^2\delta_2\right)$ \\
\hline
$g_{4,1}$    & $\frac{17 e^4}{302400 \cdot \pi^2}$ & $\frac{e}{302400 \cdot \pi^2} \left(41e^3 + 300e^2\delta_1 + 270e\delta_1^2 - 600\delta_1^3\right)$ & $-\frac{e}{100800 \cdot \pi^2} \left(83e^3 + 290e^2\delta_1 + 270e^2\delta_2\right)$ \\
\hline
$g_{4,2}$   & $\frac{e^4}{120960 \cdot \pi^2}$ & $\frac{e}{60480 \cdot \pi^2} \left(5e^3 + 24e^2\delta_1 + 72e\delta_1^2 + 96\delta_1^3\right)$ & $\frac{e}{40320 \cdot \pi^2} \left(69e^3 + 304e^2\delta_1 + 336e^2\delta_2\right)$ \\
\hline
\end{tabular}
\caption{Higher-order Wilson coefficients in EHEFT defined in \cref{eq:higher order Wilson EH} obtained by matching with a more general theory containing a massive spinning particle of spin $S$ interacting non-minimally with multipole coefficients $\delta_i$.}
\label{tab:higher order coefficients EH}
\end{table}
\renewcommand{\arraystretch}{1.25}
\begin{table}[htb]
\centering
\begin{tabular}{|c|c|c|c|c|c|}
\hline
 & \( S=0 \) & \( S=\frac{1}{2} \) & \( S=1 \) & \( S=\frac{3}{2} \) & \( S=2 \) \\
\hline
$a_{2,0}$ & \(\frac{3}{16 \cdot 280280 \cdot \pi^2}\) & \(\frac{1}{16 \cdot 30576 \cdot \pi^2}\) & \(\frac{1009 + 11011 \cdot \delta_2}{16 \cdot 9459450 \cdot \pi^2}\) & \(\frac{13571 - 100100 \cdot \delta_2}{16 \cdot 25225200 \cdot \pi^2}\) & \(\frac{564792 + 2871869 \cdot \delta_2}{16 \cdot 75675600 \cdot \pi^2}\) \\
\hline
$a_{2,1}$ & \(\frac{1}{16 \cdot 1081080 \cdot \pi^2}\) & \(\frac{29}{16 \cdot 5405400 \cdot \pi^2}\) & \(\frac{433 + 2860 \cdot \delta_2}{16 \cdot 10810800 \cdot \pi^2}\) & \(\frac{503 - 2860 \cdot \delta_2}{16 \cdot 1081080 \cdot \pi^2}\) & \(\frac{25009 + 118690 \cdot \delta_2}{16 \cdot 2162160 \cdot \pi^2}\) \\
\hline
$a_{4,0}$ & \(\frac{127}{4410806400 \cdot \pi^2}\) & \(\frac{2813}{35286451200 \cdot \pi^2}\) & \(\frac{2249 + 28050 \cdot \delta_2}{10291881600 \cdot \pi^2}\) & \(\frac{114551 - 938400 \cdot \delta_2}{123502579200 \cdot \pi^2}\) & \(\frac{90853 + 453016 \cdot \delta_2}{6175128960 \cdot \pi^2}\) \\
\hline
$a_{4,1}$ & \(\frac{1}{653452800 \cdot \pi^2}\) & \(\frac{19}{1871251200 \cdot \pi^2}\) & \(\frac{3963 + 23800 \cdot \delta_2}{41167526400 \cdot \pi^2}\) & \(\frac{15577 - 80920 \cdot \delta_2}{10291881600 \cdot \pi^2}\) & \(\frac{418909 + 1897812 \cdot \delta_2}{8233505280 \cdot \pi^2}\) \\
\hline
$a_{4,2}$ & \(\frac{29}{13722508800 \cdot \pi^2}\) & \(\frac{19}{1286485200 \cdot \pi^2}\) & \(\frac{5939 + 35224 \cdot \delta_2}{41167526400 \cdot \pi^2}\) & \(\frac{46933 - 243236 \cdot \delta_2}{20583763200 \cdot \pi^2}\) & \(\frac{3139961 + 14231924 \cdot \delta_2}{41167526400 \cdot \pi^2}\) \\
\hline
$a_{6,0}$ & \(\frac{331}{245828943360 \cdot \pi^2}\) & \(\frac{227}{64691827200 \cdot \pi^2}\) & \(\frac{31681 + 433048 \cdot \delta_2}{3687434150400 \cdot \pi^2}\) & \(\frac{713133 - 6361124 \cdot \delta_2}{22124604902400 \cdot \pi^2}\) & \(\frac{12964302 + 63328463 \cdot \delta_2}{22124604902400 \cdot \pi^2}\) \\
\hline
$a_{6,1}$ & \(\frac{1}{22348085760 \cdot \pi^2}\) & \(\frac{1}{3047466240 \cdot \pi^2}\) & \(\frac{2561 + 14364 \cdot \delta_2}{670442572800 \cdot \pi^2}\) & \(\frac{25451 - 125172 \cdot \delta_2}{335221286400 \cdot \pi^2}\) & \(\frac{2128117 + 9399186 \cdot \delta_2}{670442572800 \cdot \pi^2}\) \\
\hline
$a_{6,2}$ & \(\frac{1}{12033584640 \cdot \pi^2}\) & \(\frac{677}{938619601920 \cdot \pi^2}\) & \(\frac{43163 + 234612 \cdot \delta_2}{4693098009600 \cdot \pi^2}\) & \(\frac{11605 - 56658 \cdot \delta_2}{61751289600 \cdot \pi^2}\) & \(\frac{37100263 + 163721670 \cdot \delta_2}{4693098009600 \cdot \pi^2}\) \\
\hline
$a_{6,3}$ & \(\frac{1}{11587896320 \cdot \pi^2}\) & \(\frac{23}{26072766720 \cdot \pi^2}\) & \(\frac{2687 + 14288 \cdot \delta_2}{223480857600 \cdot \pi^2}\) & \(\frac{367 - 1784 \cdot \delta_2}{1470268800 \cdot \pi^2}\) & \(\frac{16455419 + 72593528 \cdot \delta_2}{1564366003200 \cdot \pi^2}\) \\
\hline
\end{tabular}
\caption{
Higher-order Wilson coefficients in GREFT defined in \cref{eq:higher order Wilson GREFT} obtained matching with a more general theory containing a massive spinning particle of spin $S$ with an anomalous gravi-magnetic quadrupole moment parametrized by $\delta_2$. 
}
\label{tab:higher order coefficients GREFT}
\end{table}

\newpage
\bibliographystyle{JHEP}
\bibliography{greft}

\end{document}